%% file: 4lqcdew.tex
\documentclass[a4paper]{article} \pdfoutput=1 \usepackage[T1]{fontenc}
\usepackage{graphicx} \usepackage{amsmath} \usepackage{xspace}
\usepackage[numbers,sort&compress]{natbib} \usepackage{subfig}
\usepackage[shortlabels]{enumitem} \usepackage{pdflscape}
\usepackage{pstricks}
\usepackage{graphicx}
\usepackage{xspace}
\usepackage{array}
\usepackage{multirow}
\usepackage{afterpage}
\usepackage[mathscr]{euscript}
\usepackage{ulem}
\usepackage{jheppub}

\hypersetup{
  pdfauthor={S. Kallweit, M. Grazzini, J. M. Lindert, S. Pozzorini, M. Wiesemann},
  pdftitle={NNLO QCD+NLO EW with MATRIX+OpenLoops: precise predictions for vector-boson pair production}
}

\input{defs}

\preprint{
  \begin{flushright}
    MPP-2019-175\\
    ZU-TH 49/19
  \end{flushright}
}

\author[a]{M.~Grazzini,}
\author[b]{S.~Kallweit,}
\author[c]{J.~M.~Lindert,}
\author[a]{S.~Pozzorini,}
\author[d]{and M.~Wiesemann}

\affiliation[a]{Physik-Institut, Universit\"at Z\"urich,
		Winterthurerstrasse 190, 
		CH-8057 Z\"urich,
		Switzerland }

\affiliation[b]{Dipartimento di Fisica, Universit\`a degli Studi di Milano-Bicocca and
INFN, Sezione di Milano-Bicocca, I-20126, Milan, Italy
		}

\affiliation[c]{Department of Physics and Astronomy, University of Sussex, Brighton BN1 9QH, UK}

\affiliation[d]{Max-Planck-Institut f\"ur Physik, D-80805 Munich, Germany}

\emailAdd{grazzini@physik.uzh.ch}
\emailAdd{stefan.kallweit@cern.ch}
\emailAdd{j.lindert@sussex.ac.uk}
\emailAdd{pozzorin@physik.uzh.ch}
\emailAdd{wieseman@mpp.mpg.de}

\title{NNLO QCD\,+\,NLO EW with M{\large ATRIX}+O{\large PEN}L{\large OOPS}: precise predictions for vector-boson pair production}

\abstract{%
We present the first combination of NNLO QCD and NLO EW 
corrections for vector-boson pair production at the LHC.
We consider all final states with two, three and four charged
leptons, including resonant and non-resonant diagrams,
spin correlations and off-shell effects.
Detailed predictions are discussed for three representative channels
corresponding to \wpwm, \wpmz and \zz production.
Both QCD and EW corrections are very significant, and
the details of their combination can play a crucial role 
to achieve the level of precision demanded by experimental analyses.
In this context we point out nontrivial issues that arise at large transverse
momenta, where the EW corrections are strongly enhanced by Sudakov
logarithms and the QCD corrections can feature so-called giant
$K$-factors.
Our calculations have been carried out 
in the \MatrixOpenLoops framework and 
can be extended to the production of an arbitrary 
colour singlet 
in hadronic collisions, provided that the 
required two-loop QCD amplitudes are available.
Combined NNLO QCD and NLO EW predictions 
for the full set of massive diboson processes 
will be made publicly available in the next release of \Matrix{}{}
and will be instrumental in advancing precision diboson studies
and new-physics searches at the LHC and future hadron colliders.

}


\begin{document}

\maketitle
\flushbottom

\section{Introduction}
\label{sec:intro}

The production of a pair of massive vector bosons, \wpwm, \wpmz and \zz,
plays a crucial role in various areas of the LHC physics programme.
Experimental studies for this class of processes 
lead to precise tests of the Standard Model (SM) 
at energies that range from the electroweak (\EW{}) scale up to the 
\TeV{} regime.
In particular, due to the high sensitivity to anomalous 
trilinear couplings, differential measurements at high 
transverse  momenta  probe the gauge symmetry structure of \EW{} 
interactions and provide an excellent chance to detect 
indirect effects of physics beyond the Standard Model (BSM).
Diboson final states are widely studied
also in the context of direct BSM searches and in 
Higgs-boson measurements, where they are non-trivial irreducible backgrounds, as, for example, in the \mbox{$\PH\to \wpwm$} and \mbox{$\PH\to\zz$} analyses~\cite{Sirunyan:2019twz,Aaboud:2018jqu,Aaboud:2018puo,Aaboud:2018wps,Sirunyan:2018koj,Sirunyan:2018vkx,Sirunyan:2018sgc,Aaboud:2017oem}. 
The increasing level of precision of experimental measurements 
calls for continuous
improvements 
in the theoretical description of diboson production at the LHC. In particular, SM measurements of 
vector-boson pair production~\cite{Aaboud:2019nkz,Aaboud:2019lgy,Aaboud:2019lxo,Aaboud:2019gxl,Sirunyan:2019bez,Aaboud:2017qkn,Aaboud:2017rwm,Sirunyan:2017zjc,Aaboud:2016yus,Khachatryan:2016tgp} are approaching percent-level precision, which can be matched on the 
theoretical side only by highest-order computations in both QCD and EW perturbation theory.

Leptonically decaying vector-boson pairs yield clean experimental signatures with charged leptons and missing energy. 
Theoretical predictions for on-shell \wpwm \cite{Gehrmann:2014fva}, \wpmz \cite{Grazzini:2016swo} and \zz \cite{Cascioli:2014yka,Heinrich:2017bvg}  production, and their full leptonic final states \cite{Grazzini:2015hta,Grazzini:2016ctr,Grazzini:2017ckn,Kallweit:2018nyv}, are available up to
next-to-next-to-leading order (NNLO) in
QCD in the \Matrix framework~\cite{Grazzini:2017mhc}.
The effect of the resummation of the large logarithmic terms at small transverse momenta up to next-to-next-to-leading logarithmic accuracy (NNLL) has been included in \citere{Grazzini:2015wpa} in the same framework, while the matching of the NNLO calculation with parton showers has been
presented in \citere{Re:2018vac}.
For charge-neutral final states, loop-induced contributions from gluon fusion play an important role, and they are known at ${\cal O}(\as^3)$~\cite{Caola:2015psa,Caola:2015rqy,Grazzini:2018owa}.

In order to reach the level of precision required by present and future experimental analyses, higher-order QCD predictions need to be supplemented by
the effect of EW corrections.
In general, the dominant EW corrections arise from QED radiation effects in distributions of the final-state leptons, and from large Sudakov logarithms in the high-energy tails of kinematic distributions~\cite{Denner:2000jv,Accomando:2004de}.
Calculations of diboson production at NLO EW have first been carried out
for stable vector
bosons~\cite{Bierweiler:2012kw,Bierweiler:2013dja,Baglio:2013toa,Gieseke:2014gka},
and subsequently extended to the case of off-shell vector bosons 
decaying into any final state with 
four~\cite{Biedermann:2016yvs,Biedermann:2016lvg} and three charged
leptons~\cite{Biedermann:2017oae}, \ie $\ell\ell\ell'\ell'$ and
$\ell\ell\ell'\nu_{\ell'}$ with \mbox{$\ell=\ell'$} or \mbox{$\ell\neq \ell'$},
as well as final states with two charged leptons of different flavour plus missing
energy~\cite{Biedermann:2016guo}, 
\ie $\ell\nu_{\ell}\ell'\nu_{\ell'}$ with \mbox{$\ell\neq \ell'$}. 
Results at NLO \QCDpEW accuracy for any final state with two charged
leptons plus missing energy, \ie $\ell\ell\nu_{\ell'}\nu_{\ell'}$ and $\ell\nu_{\ell}\ell'\nu_{\ell'}$
with \mbox{$\ell=\ell'$} or \mbox{$\ell\neq \ell'$},
have been presented in \citere{Kallweit:2017khh}.
For the case of on-shell \wpwm production, EW two-loop Sudakov corrections have been studied up to \NNLL accuracy~\cite{Kuhn:2011mh}.

In this paper, we present the first combination of NNLO QCD and NLO EW corrections for massive vector-boson pair production processes in their leptonic decay modes at the LHC. 
We consider all the \wpwm, \wpmz and \zz production processes  
with two, three and four final-state charged leptons.
Spin correlations, interferences and off-shell effects are 
included throughout. While QCD corrections have no effect on leptonic
vector-boson decays, we note that the NLO EW corrections affect both the production
of the virtual vector-boson pairs and their subsequent decays. Moreover, the 
off-shell \mbox{$pp\to 4$\,fermion} processes at hand involve also 
irreducible NLO EW effects 
that cannot be separately attributed to vector-boson pair production or decay.

The combination of QCD and EW corrections is discussed in detail.
In particular, by comparing different combination prescriptions 
we  gain insights into theoretical uncertainties associated with 
higher-order effects beyond NNLO QCD+NLO EW.
In this context we devote particular attention to 
the tails of various kinematic distributions
that feature giant QCD $K$-factors~\cite{Rubin:2010xp,Baglio:2013toa} 
together with large Sudakov-type EW corrections, 
resulting  in sizeable higher-order uncertainties.
As we will show, such uncertainties can be 
tamed by requiring that the two 
vector bosons are comparably hard in the high-$\pT$ regime.
This can be achieved either via
direct selection cuts on the vector bosons or through a jet veto.

The calculations presented in this paper have been carried out using \MatrixOpenLoops.
Our implementation of the NNLO QCD and NLO EW corrections is completely general, and can be applied to the production of an arbitrary colour singlet in hadronic collisions,
provided that the required two-loop QCD amplitudes are available. This paves the way to extend the various process libraries that are available
in the current public version of \Matrix~\cite{Grazzini:2017mhc} to NNLO\,QCD+NLO\,EW accuracy.

The paper is organised as follows. In   \sct{se:anatomy} we outline the calculation: We describe the employed methods and tools (\sct{sec:tools}), the leptonic channels contributing to vector-boson pair processes (\sct{sec:channels}) and the diagrams contributing to their QCD (\sct{sec:QCD}) and EW (\sct{sec:ew}) corrections. In \sct{sec:giantK} we discuss the phenomenologically important issue of giant
$K$-factors, and in
\sct{sec:combination} we present different procedures to combine QCD and EW corrections.
Our numerical results for vector-boson pair production processes are presented in \sct{sec:pheno}, in the inclusive case (\sct{sec:results_inclusive}) and with a veto against hard QCD radiation applied (\sct{sec:results_veto}).
Finally, in \sct{sec:summa} we draw our conclusions.

\section{Vector-boson pair production in \NNLO \texorpdfstring{$\boldsymbol{\QCD\!}$}{QCD} 
and \NLO \texorpdfstring{$\boldsymbol{\EW\!}$}{EW}}
\label{se:anatomy}

\subsection{NNLO QCD+EW predictions with \MatrixOpenLoops}
\label{sec:tools}

The NNLO QCD and NLO EW computations for vector-boson pair production processes presented 
in this paper are carried out within the \MatrixOpenLoops framework.
The core of the \Matrix{} framework \cite{Grazzini:2017mhc} is the Monte Carlo program \Munich{}\footnote{\Munich{} is the 
abbreviation of ``MUlti-chaNnel Integrator at Swiss~(CH) precision'' --- an automated parton-level
NLO generator by S.~Kallweit.}
which includes a fully automated implementation of the dipole-subtraction method
at NLO QCD~\cite{Catani:1996jh,Catani:1996vz} and NLO EW~\cite{Dittmaier:1999mb,Dittmaier:2008md,Gehrmann:2010ry,Kallweit:2017khh,Schonherr:2017qcj},
combined with an efficient multi-channel integration algorithm.
The required tree-level and one-loop
amplitudes, including spin and colour correlations for the subtraction of
infrared divergences, are calculated with {\sc
OpenLoops~2}~\cite{Buccioni:2019sur}. 
This new version of the \OpenLoops program supports the automated generation
of matrix elements for any SM process at NLO QCD+NLO EW. Scattering
amplitudes are built with a new numerical recursion that combines the original method
of \citere{Cascioli:2011va} with the on-the-fly reduction technique of~\citere{Buccioni:2017yxi}.
To avoid numerical instabilities in critical phase space regions,
this new algorithm is equipped with a sophisticated system that 
automatically detects the occurrence of spurious singularities in 
the reduction identities and circumvents them by using algebraic tricks in combination
with analytic expansions up to any order in rank-two Gram determinants~\cite{Buccioni:2017yxi,Buccioni:2019sur}. 
Residual instabilities are avoided in a systematic way by means of an
efficient hybrid-precision system, which carries out certain critical operations 
in quadruple precision while using double precision for the bulk of the 
calculation.~\cite{Buccioni:2019sur}. 
These new techniques improve the numerical stability of one-loop amplitudes 
in a very significant way, especially in the regions where one parton becomes
soft or collinear.
In {\sc OpenLoops~2} scalar integrals from {\sc Collier}~\cite{Denner:2016kdg} 
and {\sc OneLOop}~\cite{vanHameren:2010cp} are used. 
Our NLO EW results for vector-boson pair processes have been validated systematically, starting
  from an in-detail comparison of all results in~\citere{Kallweit:2017khh} between \Sherpa and \Munich{}.
All relevant EW one-loop amplitudes were validated between \OpenLoops and \RecolaTwo~\cite{Denner:2017wsf} at the level of phase space points.
  Further, in \citere{Bendavid:2018nar} NLO EW amplitudes and cross sections  for $pp \to \ell^+ \ell^- \ell'^+ \ell'^-$
  and $pp \to \ell^+ \ell'^- \nu_{\ell} \bar \nu_{\ell'}$ were checked amongst various public tools~\cite{Actis:2016mpe,Hirschi:2011pa,Frederix:2018nkq,Cullen:2014yla,Honeywell:2018fcl,Schonherr:2017qcj}.

The required two-loop amplitudes for NNLO QCD calculations 
need to be implemented on a
process-by-process basis. For the calculation of off-shell vector-boson pair processes,
\Matrix uses the two-loop \mbox{$q\bar{q}\to VV'$}
amplitudes of~\citere{Gehrmann:2015ora} as provided by the {\sc VVamp}
package.
The subtraction of IR singularities at NNLO is achieved through a fully
general  implementation of the $q_T$-subtraction
formalism~\cite{Catani:2007vq}. In this context, \Matrix{} 
implements an automated extrapolation procedure to calculate integrated 
cross sections in the limit in which the cutoff parameter in the $q_T$-subtraction procedure
goes to zero~\cite{Grazzini:2017mhc}.

The $q_T$-subtraction method
 and its implementation in
\Matrix were originally limited to colour-singlet final states, but have
been recently extended to provide NNLO QCD predictions for top-quark pair
production as well~\cite{Catani:2019iny,Catani:2019hip}.
Despite the fact that in this paper we focus on massive vector boson pairs,
our implementation is completely general, and can 
directly be used for any other processes with a colourless final state, e.g.~\mbox{$pp\to VH$}, $VVV$, etc., 
where $V$ refers to either on-shell vector bosons,
\mbox{$V=\{Z,W^{\pm}\}$}, or off-shell leptonic final states,
\mbox{$V=\{\ell^+\ell^-, \nu\bar \nu,\ell^-\bar\nu,\ell^+\nu\}$}.  
The only restriction
is given by the availability of the corresponding two-loop amplitudes.

\renewcommand\arraystretch{1.5}
\begin{table}
\begin{center}
\begin{tabular}{c|c|c|c}
&&contributing & shorthand\\[-3mm]
acronym     & process                                & resonances  &  in
this paper \\\hline
4l-SF-ZZ    & $p p \to \ell^+ \ell^- \ell^+ \ell^- $                &  ZZ &
\\
4l-DF-ZZ    & $p p \to  \ell^+ \ell^- \ell'^+ \ell'^-$              &  ZZ &
\\
3l-SF-WZ    & $p p  \to \ell^+ \ell^-  \ell \nu_{\ell}  $           &  WZ &      
\\
3l-DF-WZ    & $p p  \to \ell^+ \ell^- \ell' \nu_{\ell'} $           &  WZ  &
WZ    
\\
2l-SF-ZZ    & $p p \to  \ell^+ \ell^- \nu_{\ell'} \bar \nu_{\ell'} $&  ZZ & ZZ
\\
2l-SF-ZZWW  & $p p \to  \ell^+ \ell^- \nu_{\ell} \bar \nu_{\ell} $  &  ZZ,WW &
\\
2l-DF-WW    & $p p  \to \ell^+ \ell'^- \nu_{\ell} \bar \nu_{\ell'} $&  WW & WW    
\\
\end{tabular}
\end{center}
\caption{Complete list of diboson processes that are implemented in \Matrix
and will be upgraded to NNLO
QCD+NLO EW accuracy in the forthcoming code release. The last column indicates the shorthands used for the
three representative processes presented in this paper.
In this table it is implicitly understood that \mbox{$\ell'\neq \ell$}.}
\label{tab:process_list}
\end{table}

\subsection{Overview of leptonic channels in vector-boson pair production}
\label{sec:channels}

The present version of \Matrix implements the full set of \mbox{$pp\to4$\,fermion}
processes that involve \wpwm, \wpmz or \zz resonances
in all channels with final states containing two, three or four charged
leptons.\footnote{Diboson final states with less than two charged leptons
plus missing energy are only marginally relevant from the phenomenological point of view.}
The complete set of processes and corresponding acronyms used in this paper are listed in
\refta{tab:process_list}.
The various final states are categorised according to the number of charged
leptons and further split into different-flavour~(DF) and same-flavour~(SF) 
channels according to the flavours of charged leptons.
Note that the channels 2l-SF-ZZ and 2l-SF-ZZWW yield identical experimental signatures 
and should thus be combined for phenomenological applications.\footnote{For the 2l-SF-ZZWW channel it was shown that 
the interference between topologies with \ww and \zz resonances 
can be safely neglected both at NLO EW~\cite{Kallweit:2017khh} and at NNLO
QCD~\cite{Kallweit:2018nyv}. Thus, the process 2l-SF-ZZWW, with
\zz--\ww interferences, can be approximated by the incoherent superposition of the
2l-SF-ZZ and 2l-DF-WW processes~\cite{Kallweit:2017khh,Kallweit:2018nyv}.
}

We have combined NNLO~QCD with NLO~EW corrections within the \MatrixOpenLoops
framework for all processes listed in \refta{tab:process_list}. 
This implementation will be made publicly available with the forthcoming version of \Matrix{},
and in this
paper we illustrate the effect of NNLO~QCD+ NLO~EW corrections on the representative
channels 2l-SF-ZZ, 2l-DF-WW and 3l-DF-WZ. For brevity, we will refer to
these three channels as \zz, \ww and \wz production, respectively.
As pointed out in the introduction, all relevant \mbox{$pp\to4$\,lepton}
matrix elements are computed exactly, \ie 
without applying any resonance approximation.
All Feynman diagrams with double-, single- and non-resonant
topologies are consistently included
at each perturbative order using the complex-mass scheme~\cite{Denner:2005fg}.
Therefore off-shell effects, interferences and spin correlations are fully taken into account
throughout.

\begin{figure}[t]
\begin{center}
\begin{tabular}{ccccccc}
\includegraphics[width=.20\textwidth]{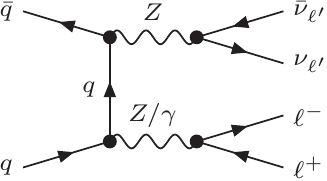} & &
\includegraphics[width=.20\textwidth]{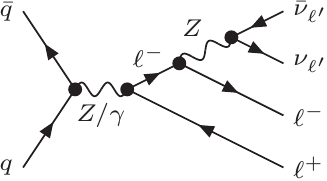} & &
\includegraphics[width=.20\textwidth]{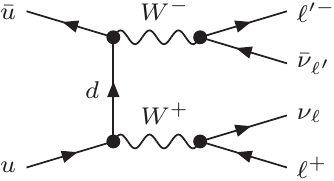} & &
\includegraphics[width=.20\textwidth]{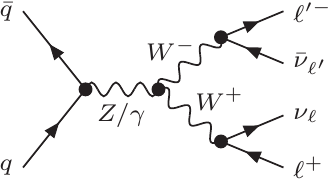} \\[0ex]
(a) & & (b) & & (c) & & (d) \\[2ex]
\includegraphics[width=.20\textwidth]{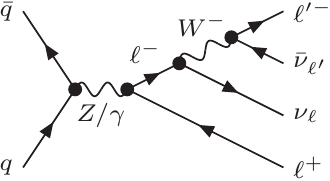} & &
\includegraphics[width=.20\textwidth]{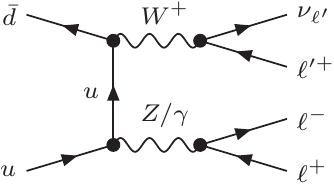} & &
\includegraphics[width=.20\textwidth]{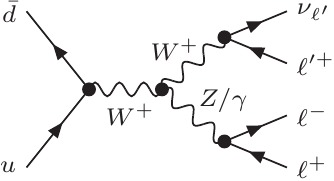} & &
\includegraphics[width=.20\textwidth]{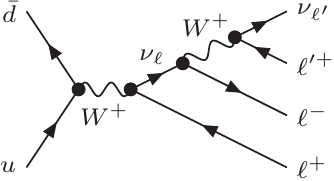} \\[0ex]
(e) & & (f) & & (g) & & (h)
\end{tabular}
\end{center}\vspace{-0.4cm}
\caption[]{
  \label{fig:LOdiagrams}
 Sample LO diagrams for 2l-SF-ZZ (a-b), 2l-DF-WW (c-e), and 3l-DF-WZ (f-h).}
\end{figure}

\begin{figure}[t]
\vspace{0.5cm}
\begin{center}
\begin{tabular}{ccccccc}
\includegraphics[width=.20\textwidth]{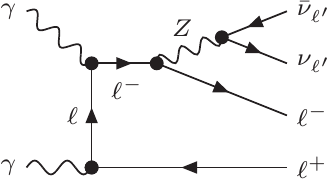} & &
\includegraphics[width=.20\textwidth]{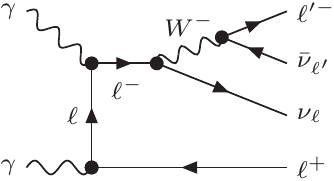} & &
\includegraphics[width=.20\textwidth]{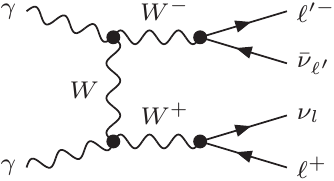} & &
\includegraphics[width=.20\textwidth]{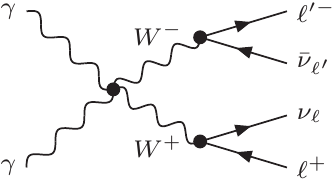} \\[0ex]
(a) & & (b) & & (c) & & (d)
\end{tabular}
\end{center}\vspace{-0.4cm}
\caption[]{
  \label{fig:LOphotons}
 Sample photon-induced LO diagrams for 2l-SF-ZZ (a) and 2l-DF-WW (b-d). There is no photon-induced LO contribution to 3l-DF-WZ.}
\end{figure}

In \fig{fig:LOdiagrams} we show representative LO Feynman diagrams 
for the selected \zz, \ww and \wz production processes. 
As illustrated in \fig{fig:LOphotons},  diboson processes with 
charge-neutral final states, \ie \zz and
\ww production, involve additional photon-induced channels.
In \MatrixOpenLoops the photon distribution function is 
treated on the same footing as the QCD parton densities.
Thus, photon-induced channels enter at the same
perturbative order as the usual $q\bar q$ channels,
and both channels are supplemented by NLO EW corrections.
This is important for a reliable description of 
certain phase space regions where photon-induced effects can be significantly 
enhanced by the opening of quark--photon channels at NLO EW.

\subsection{Higher-order QCD corrections}
\label{sec:QCD}

For vector-boson pair production processes, higher-order QCD corrections have a 
sizeable impact. The NLO QCD corrections
increase inclusive cross sections by \percentrange{40}{50} for \zz and \ww production and around
\percentrange{70}{80} for \wz production~\cite{Ohnemus:1990za,Ohnemus:1991kk,Ohnemus:1991gb,Frixione:1993yp,
Ohnemus:1994ff,Campbell:1999ah,Campbell:2011bn,Melia:2011tj,Baglio:2013toa}. The large NLO effect for \wz production originates from an approximate \textit{radiation zero} appearing in the leading helicity amplitude for \wz production at LO~\cite{Mikaelian:1979nr}, which is not present at higher orders.
Also NNLO QCD corrections have a quite
significant impact, at the level of \percent{10} or more, on the various diboson
production processes~\cite{Grazzini:2013bna,Grazzini:2015nwa,Cascioli:2014yka,Grazzini:2015hta,Gehrmann:2014fva,Grazzini:2016ctr,Grazzini:2016swo,Grazzini:2017ckn,Kallweit:2018nyv}. 

\begin{figure}
\begin{center}
\begin{tabular}{ccccccc}
\multicolumn{3}{c}{\underline{NLO QCD: virtual}} & &\multicolumn{3}{c}{\underline{NLO QCD: real}}\\[2ex]
\includegraphics[width=.20\textwidth]{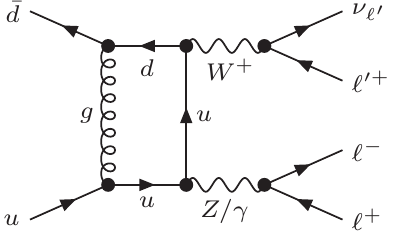} & &
\includegraphics[width=.20\textwidth]{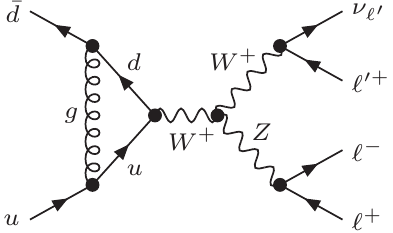} & &
\includegraphics[width=.20\textwidth]{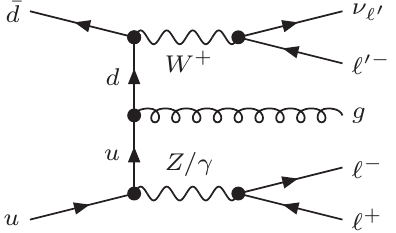} & &
\includegraphics[width=.20\textwidth]{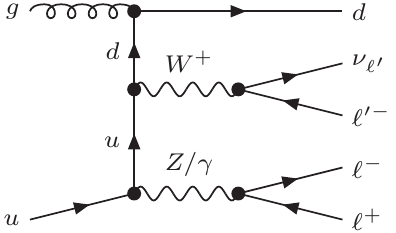} \\[0ex]
(a) & & (b) & & (c) & & (d) \\[2ex]
\multicolumn{3}{c}{\underline{NNLO QCD: double-virtual}} & &\multicolumn{3}{c}{\underline{NNLO QCD: real--virtual}}\\[2ex]
\includegraphics[width=.20\textwidth]{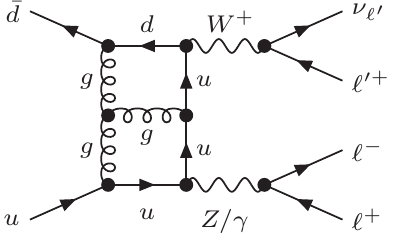} & &
\includegraphics[width=.20\textwidth]{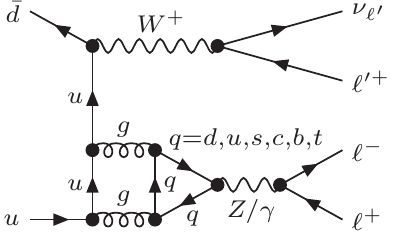}& &
\includegraphics[width=.20\textwidth]{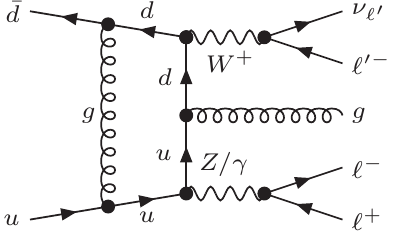} & &
\includegraphics[width=.20\textwidth]{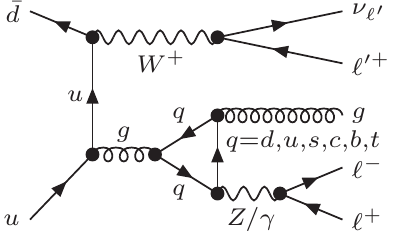} \\[0ex]
(e) & & (f) & & (g) & & (h)\\[2ex]
\multicolumn{7}{c}{\underline{NNLO QCD: double-real}}\\[2ex]
\includegraphics[width=.20\textwidth]{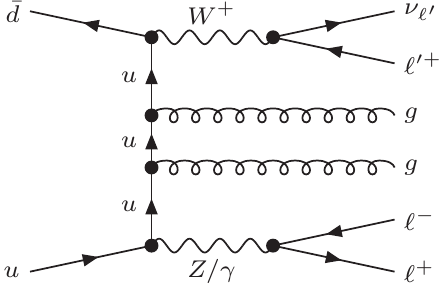} & &
\includegraphics[width=.20\textwidth]{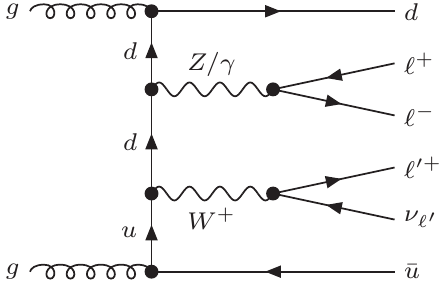} & &
\includegraphics[width=.20\textwidth]{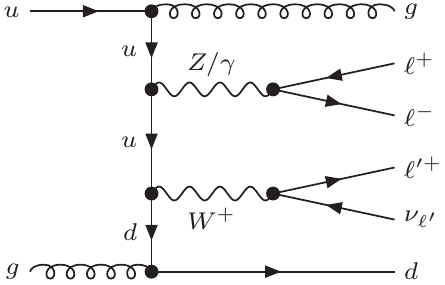} & &
\includegraphics[width=.20\textwidth]{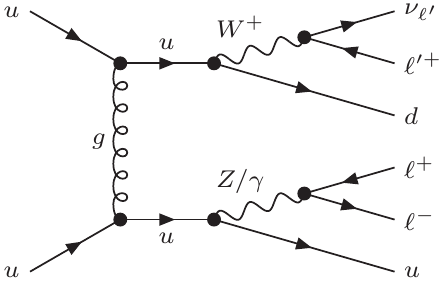}\\[0ex]
(i) & & (j) & & (k) & & (l)
\end{tabular}
\end{center}\vspace{-0.4cm}
\caption[]{
  \label{fig:QCDdiagramsWZ}
  Sample higher-order QCD diagrams for $\ell^+ \ell^- \ell'
\nu_{\ell'}(\wz)$ production:
NLO QCD diagrams of  virtual (a--b) and real (c--d) type, and 
NNLO QCD diagrams of double-virtual (e--f), real--virtual (g--h) and double-real (i--l) 
type.} 
\end{figure}

Predictions at NLO QCD require the calculation of 
virtual and real-emission matrix elements, 
while NNLO QCD corrections involve 
double-virtual, real-virtual, and
double-real  contributions.
Representative Feynman diagrams are displayed in \reffi{fig:QCDdiagramsWZ}
for the case of \wpz production.
Similar diagrams contribute also to the other diboson processes.
Only for \zz production diagrams with triple vector-boson couplings are absent.
In addition to the contributions illustrated in \reffi{fig:QCDdiagramsWZ},
\ww and \zz production involve also a loop-induced gluon-fusion channel
that enters at $\ord(\as^2)$, \ie it is part of the NNLO~QCD corrections.
The contribution of this \mbox{$gg\to VV$} channel to charge-neutral final states
is quite sizeable. It has been computed to one order higher in perturbation theory~\cite{vonManteuffel:2015msa,Caola:2015psa,Caola:2015ila,Caola:2016trd,Alioli:2016xab,Grazzini:2018owa,Grazzini:2019xxx}, 
which is assumed to be the dominant $\ord(\alphaS^3)$
correction to these processes.
In the combination of NNLO QCD and NLO EW corrections
presented in this paper, the \mbox{$gg\to VV$} channels are included 
at $\ord(\alphaS^2)$ as part of the NNLO QCD corrections, \ie neglecting $\ord(\alphaS^3)$ effects.

\subsection{Higher-order EW corrections}
\label{sec:ew}

The impact of NLO EW effects on inclusive cross sections is typically at
the few-percent level and thus important in the context of high-precision
studies. In kinematic distributions, EW corrections can be more sizeable.
In particular, in the tails of distributions that probe high-energy
scales \mbox{$Q\gg \MW$}, the EW corrections are enhanced by Sudakov logarithms~\cite{Denner:2000jv,Accomando:2004de} 
of the form 
$\alphaW \log^2\left(Q^2/\MW^2\right)$, where
\mbox{$\alphaW=g_\rw^2/(4\pi)$} denotes the SU(2) coupling strength. 
The size of EW Sudakov effects depends on $Q$ 
as well as on the SU(2)$\times$U(1) quantum numbers of the 
scattering particles. Such logarithmic effects are most pronounced 
in processes with (multiple) transversely polarised $W$ and $Z$ bosons,
and in the case of \zz, \wz and \ww production they can lead to NLO~EW 
corrections of several tens of percent at the TeV scale.

EW corrections have been studied for various vector-boson pair
production modes, treating the vector bosons
on-shell~\cite{Bierweiler:2012kw,Baglio:2013toa,Gieseke:2014gka}, and more
recently also fully
off-shell~\cite{Biedermann:2016yvs,Biedermann:2016lvg,Biedermann:2016guo,Kallweit:2017khh,Biedermann:2017oae}. 
Including off-shell and non-resonant effects is preferable, and they can play an especially important role 
in the tails of kinematic distributions~\cite{Biedermann:2016guo}. 

\begin{figure}
\begin{center}
\begin{tabular}{ccccccc}
\multicolumn{7}{c}{\underline{NLO EW: virtual}}\\[2ex]
\includegraphics[width=.20\textwidth]{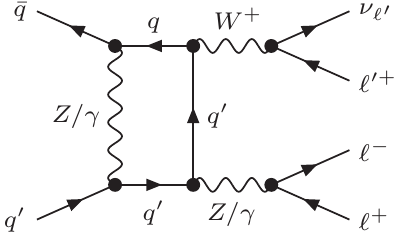} & &
\includegraphics[width=.20\textwidth]{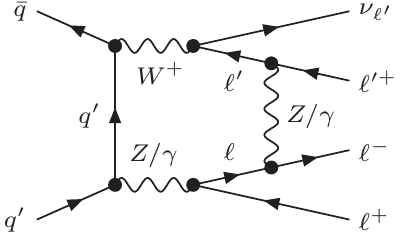} & &
\includegraphics[width=.20\textwidth]{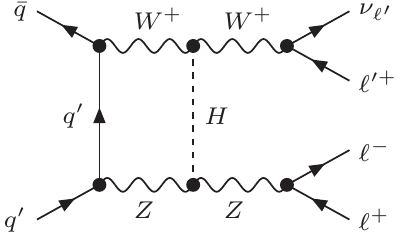} & &
\includegraphics[width=.20\textwidth]{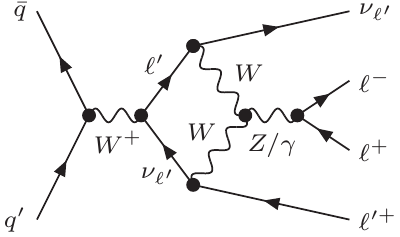} \\[0ex]
(a) & & (b) & & (c) & & (d) \\[2ex]
\includegraphics[width=.20\textwidth]{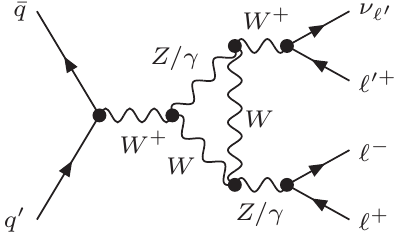} & &
\includegraphics[width=.20\textwidth]{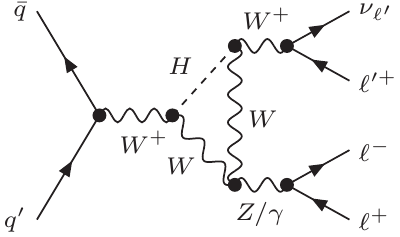} & &
\includegraphics[width=.20\textwidth]{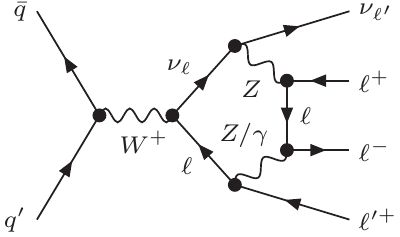} & &
\includegraphics[width=.20\textwidth]{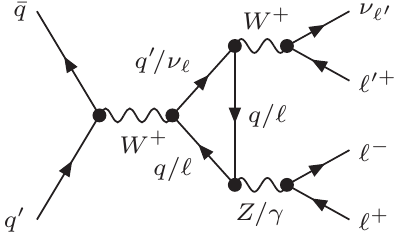} \\[0ex]
(e) & & (f) & & (g) & & (h)\\[2ex]
\multicolumn{7}{c}{\underline{NLO EW: real}}\\[2ex]
\includegraphics[width=.20\textwidth]{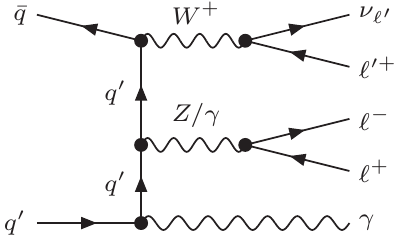} & &
\includegraphics[width=.20\textwidth]{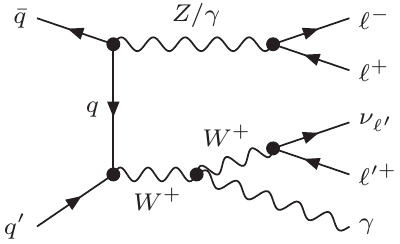} & &
\includegraphics[width=.20\textwidth]{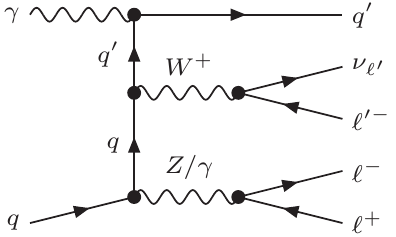} & &
\includegraphics[width=.20\textwidth]{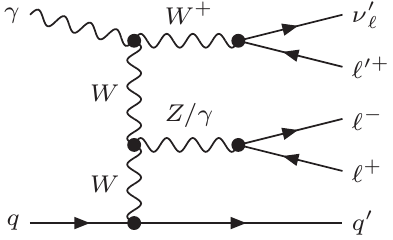} \\[0ex]
(i) & & (j) & & (k) & & (l)
\end{tabular}
\end{center}\vspace{-0.4cm}
\caption[]{
  \label{fig:EWdiagramsWZ}
  Sample NLO EW virtual (a-h) and real (i-l) diagrams for $\ell^+ \ell^- \ell' \nu_{\ell'}(\wpz)$
production.}
\end{figure}

The structure of NLO EW corrections to vector-boson pair production is
illustrated in \reffi{fig:EWdiagramsWZ}, where we show representative Feynman diagrams
for the virtual and real corrections to \wz production.\footnote{For a 
more detailed discussion of the NLO EW ingredients to off-shell \zz and \ww
production, see \citere{Kallweit:2017khh}.}
The virtual corrections enter only through the $q\bar{q}$ channel 
and involve one-loop diagrams with various combinations of photons, $Z$/$W^{\pm}$-bosons, Higgs
bosons, light fermions and heavy quarks in the loop.
Real-emission contributions consist of a $q\bar q$ channel with an
additional final-state photon, a $q\gamma$ channel with an additional
final-state quark and a corresponding $\bar q \gamma$ channel.
The extra photon couples to any external or internal charged fermion or
$W$ boson.

In the case of \zz and \ww production, the presence of additional 
\mbox{$\gamma\gamma\to VV$} channels at LO gives rise to corresponding 
virtual and real contributions at NLO EW (not shown in
\reffi{fig:EWdiagramsWZ}).
The real EW corrections to \mbox{$\gamma\gamma\to VV$} processes
with \mbox{$VV=\zz,\ww$} involve \mbox{$\gamma\gamma\to VV\gamma$} channels as well as 
\mbox{$q\gamma\to VVq$} and \mbox{$\bar q\gamma\to VV\bar q$} channels.
As discussed in \citere{Kallweit:2017khh}, 
the $q\gamma$ and $\bar q\gamma$  channels play the twofold role of 
NLO EW corrections to \mbox{$q\bar{q}\to VV$} and \mbox{$\gamma\gamma\to VV$}.
In particular, they cannot be uniquely assigned to one or the other 
LO channel. Thus, if the photon density is treated on the same 
footing as the other parton densities, both the $q\bar q$ and  $\gamma\gamma$ channels should be 
supplemented by virtual and real corrections at NLO EW.

Triboson production processes of type \mbox{$pp\to VVV$} and \mbox{$pp\to VVH$},
with \mbox{$V=W, Z$},
contribute at the same perturbative order as the NLO EW corrections to \mbox{$pp\to VV$}.
Thus, in principle, triboson production can be regarded as part of the NLO EW 
corrections to diboson production. However, diboson and triboson production yield different
experimental signatures, and are typically handled as separate processes in
experimental analyses.
For this reason, and in order to avoid
double counting between diboson and triboson production, we do not include 
\mbox{$pp\to VVV/VVH$} in the EW corrections to \mbox{$pp\to VV$}.

\subsection[Giant $K$-factors]{Giant \boldmath{$K$}-factors}
\label{sec:giantK}

\begin{figure}
\begin{center}
\begin{tabular}{ccc}
\includegraphics[width=.20\textwidth]{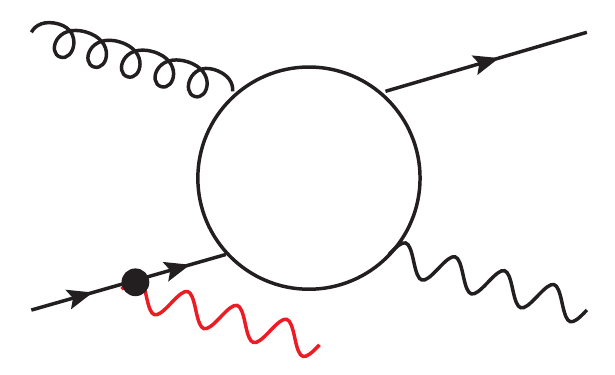}&
\includegraphics[width=.20\textwidth]{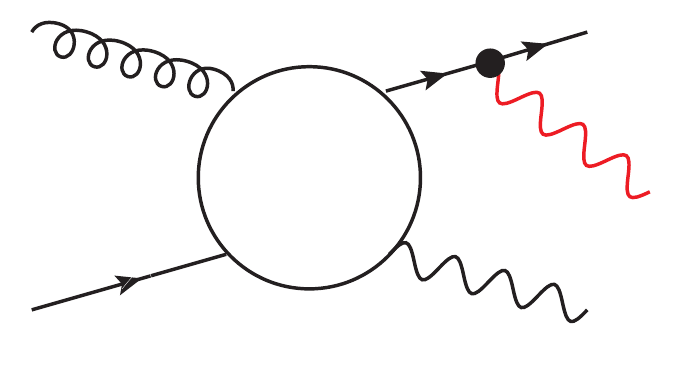}&
\includegraphics[width=.20\textwidth]{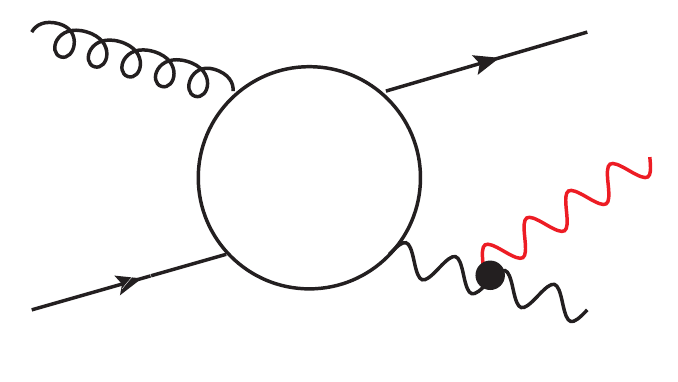}
\end{tabular}
\end{center}
\vspace{-17pt}
\caption{Generic \mbox{$pp\to VVj$} topologies and kinematic regions that 
give rise to giant $K$-factors in the quark--gluon channel at NLO QCD.
The blob denotes the hard scattering subprocess
\mbox{$g q\to V q$} at the scale \mbox{$Q\gg\MW$}, while the subleading vector boson (red) 
is radiated by one of the SU(2)$\times$U(1) charged external states
giving rise to EW logarithms of soft and collinear kind.}
\label{fig:giantKfactors}
\end{figure}

At large transverse
momenta, as shown in \refse{sec:results_inclusive}, 
the NLO QCD corrections to
vector-boson pair production can become as large as $\ord(1)$ or even 
$\ord(10)$.
These so-called giant QCD $K$-factors~\cite{Rubin:2010xp,Baglio:2013toa}
arise from phase space regions that are kinematically
forbidden at LO and are populated through the emission of
hard QCD radiation starting at NLO.
Giant $K$-factors appear in inclusive observables that require 
a vector boson with very large transverse momentum, \mbox{$p_{\rT,V_1}\sim Q \gg \MW$},
while leaving the second vector boson unconstrained. At LO, as a result of 
momentum conservation, the recoil of the first vector boson is absorbed by
the second one, \ie \mbox{$p_{\rT,V_2}=p_{\rT,V_1}$}. Instead, 
in the NLO~QCD radiative process \mbox{$pp\to VVj$}
the recoil can also be
absorbed by the additional hard jet, 
while the second vector boson becomes much softer,
\mbox{$p_{\rT,V_2}=\ord(\MW)$}. Upon integration over the phase space of the second
vector
boson, this radiative mechanism results in the cross section~\cite{Baglio:2013toa}
\beqar
\label{eq:giantkfactA}
d\sigma^{V(V)j}
\,\propto\, \sigma_{\LO}^{Vj}
\frac{\alphaW}{2\pi}\log^2 
\left(\frac{Q^2}{M_W^2}\right)\,,
\eeqar
where $\sigma_{\LO}^{Vj}$ describes the hard subprocess \mbox{$pp\to Vj$},
while the double Sudakov logarithm on the right-hand side can be understood as the inclusive probability
of radiating a soft vector boson.
Since \mbox{$\sigma_{\LO}^{Vj}/\sigma_{\LO}^{VV}\propto \alphaS/\alphaW$}, 
the radiative contribution \refwoeq{eq:giantkfactA} yields a $K$-factor to the $VV$ process proportional 
to 
\beqar
\label{eq:giantkfactB}
\frac{d\sigma^{V(V)j}}{d\sigma^{\LO}_{VV}} 
\propto\, \alphaS\,\log^2 
\left(\frac{Q^2}{M_W^2}\right)
\,\simeq\, 3\qquad\mbox{at}\quad Q=1\,\TeV\,.
\eeqar
General real-emission topologies that lead to 
giant $K$-factors are depicted in \reffi{fig:giantKfactors}.
They correspond to a hard \mbox{$pp\to Vj$} subprocess 
at the scale \mbox{$Q\gg \MW$}
supplemented by soft vector-boson
radiation. 
The corresponding kinematic regions
will  be referred
to as \hardVj regions,
and they are characterised by a hard jet
with \mbox{$p_{\rT,j}\sim Q$} and a large gap between the leading and subleading
vector boson, \mbox{$p_{\rT,V_2}\ll p_{\rT,V_1}$}.
Conversely, standard QCD radiation effects correspond to a hard subprocess
\mbox{$pp\to VV$} at the scale $Q$ and QCD radiation at scales well below $Q$.
In this case the two vector bosons are comparably hard, and such phase space
regions will be classified as hard-$VV$ regions.

Noteworthy, giant $K$-factors can also arise at NLO EW, where they appear in 
\mbox{$\gamma q\to VV q$} real-emission processes
with a hard \mbox{$\gamma q\to V q$} subprocess and soft vector-boson radiation,
as well as in crossing-related \mbox{$q\bar q\to VV \gamma$} processes with a hard \mbox{$q\bar q\to V \gamma$} subprocess.
At NLO EW, in addition to the topologies of \reffi{fig:giantKfactors}
with gluons replaced by photons, 
also extra topologies where the soft vector boson is radiated off 
external photons arise.
Here, the giant $K$-factor mechanism leads to NLO EW effects of order $\alphaW\log^2(Q^2/\MW^2)$, and
these are dominated by the \mbox{$\gamma q\to VV q$} channel. 

The appearance of giant $K$-factors at NLO raises important questions
concerning the convergence of the perturbative expansion and the combination
of QCD and EW corrections.  In this respect, it is important to note that,
contrary to QCD logarithmic effects of soft and collinear origin,
the large logarithms in \refeq{eq:giantkfactA} do not contribute to
all orders in $\alphaS$.  
In fact, such logarithms do not arise from soft QCD radiation, but from 
soft vector-boson radiation in combination with the opening of the hard
\mbox{$pp\to V(V)j$} channel at NLO QCD. Since this happens only when moving from LO to
NLO QCD, higher-order QCD corrections beyond NLO are free from further 
giant $K$-factors.\footnote{Here, we assume that in diboson production at the scale \mbox{$Q\gg \MW$} at least one
vector boson with \mbox{$p_{\rT,V_1}=\ord(Q)$} is required.  Otherwise, allowing
both vector bosons to become soft would result into giant NNLO QCD
$K$-factors of the form $\alphaS^2\log^4(Q^2/\MW^2)$ stemming from hard dijet
 topologies.}
Note also that the availability of NNLO QCD corrections makes it possible to
verify the stability of the perturbative expansion beyond NLO 
and to arrive at reliable QCD predictions for observables that
feature giant $K$-factors.

For what concerns the combination of QCD and EW corrections, the presence of
giant $K$-factors raises more serious issues. In particular, the fact that in the relevant 
high-$\pT$ regions the NLO QCD and NLO EW corrections are both strongly
enhanced implies sizeable theoretical uncertainties from 
large unknown mixed QCD--EW NNLO effects.
In principle, depending on the observable and the kinematic region, mixed QCD--EW effects
can be approximated through a factorised description of QCD and EW
corrections (see \refse{sec:combination}).
However, such a factorisation can be justified only in cases where QCD and EW corrections
are both dominated by soft corrections with respect to the same hard
subprocess.
In the case at hand, this condition is not fulfilled
since NLO EW effects are driven by logarithmic Sudakov
corrections to hard $VV$ production, whereas giant QCD $K$-factors 
are dominated by soft EW boson radiation on top of hard $Vj$ production.
Actually, the leading source of $\ord(\alphaS\alpha)$ corrections 
is given by the NLO EW corrections to 
the enhanced  \mbox{$pp\to VVj$} channel,
which cannot be captured through a 
naive factorised combination of the NLO QCD and NLO EW corrections to \mbox{$pp\to VV$}.

When presenting our results in \sct{sec:pheno}, the problem of giant $K$-factors in the inclusive phase space will be illustrated. 
We will show that giant $K$-factors can be avoided by means of selection cuts that require a similar hardness of the 
two vector bosons, \eg by direct requirements on the hardness of the softer vector boson or by imposing a veto against hard QCD radiation.
This will restrict the phase space to \hardVV{} topologies and suppress \hardVj production. 
Besides reducing the size of mixed QCD--EW higher-order effects and their respective theoretical uncertainties,
selecting \hardVV topologies enhances the sensitivity of experimental measurements that aim at 
extracting new-physics effects in vector-boson pair processes, such as anomalous triple gauge couplings, 
from the tails of kinematic distributions.
On the other hand, a reliable inclusive description of diboson production is indispensable for background simulations
in direct searches at the TeV scale. This can be achieved by merging \mbox{$pp\to VV$} and \mbox{$pp\to VVj$} production including \NLO~\QCD and \NLO~\EW corrections 
as demonstrated in~\citere{Kallweit:2015dum}. The extension of this approach to \NNLO\,\QCDpEW is beyond the scope of the present paper.

\subsection{Combination of QCD and EW corrections}
\label{sec:combination}

When QCD and EW corrections are both large, also NNLO
mixed QCD--EW effects of relative ${\cal O}(\as\alpha)$ and beyond can become important.
In order to gain insights into such higher-order effects,  
we consider a standard additive combination of 
NNLO~QCD and NLO~EW corrections and compare it 
against factorised combination prescriptions.
To this end, we express 
higher-order effects in terms of relative correction factors with respect to  
the LO differential cross section,
\begin{eqnarray}
\label{eq:LOdiff}
\sig{\LO}{} \,=\,\sig{\LO}{q\bar q}+\sig{\LO}{\gamma\gamma}\,,
\end{eqnarray}
which involves $\ord(\alpha^4)$  contributions from the
$q\bar q$ and $\gamma\gamma$ channels.\footnote{Note that the 
$\gamma\gamma$ channel contributes only to \zz and \ww production.
The same holds for the $gg$ channel contributing at NNLO QCD.}
Higher-order QCD contributions can be cast into the form 
\begin{eqnarray}
\label{eq:QCDcorr1}
\sig{\NNLO\,\QCD}{}
&=&
\sig{\LO}{}\left(1+\delt{\QCD}{}\right)
+\sig{\LO}{gg}\,,
\end{eqnarray}
where $\sig{\LO}{gg}$ is the 
$\ord(\alphaS^2\alpha^4)$ contribution of the 
loop-induced $gg$ channel, 
and all other QCD corrections are embodied in 
the correction factor $\delt{\QCD}{}$, which
includes the $\ord(\alphaS)$ and $\ord(\alphaS^2)$ corrections of the  
$q\bar q$, $q g/\bar{q}g$, $g g$ and $qq/\bar q\bar q$ channels.\footnote{Here 
and in the following, 
higher-order contributions (or terms) of $\ord(\as^n \alpha^{4+m})$
are also referred to as corrections (or effects) of $\ord(\as^n\alpha^m)$.
}
Similarly, the NLO EW cross section can be written as
\begin{eqnarray}
\sig{\NLO\,\EW}{} &=&
\sig{\LO}{}\left(1+\delt{\EW}{}\right)\,,
\end{eqnarray}
where all $\ord(\alpha)$ corrections in the 
$q\bar q$, $\gamma\gamma$ and $q\gamma$ (including $\bar q \gamma$ is implicitly understood) 
channels are incorporated into the factor $\delt{\EW}{}$.
For the combination of QCD and EW corrections we consider three different
prescriptions.
\paragraph{\NNLO\,\QCDpEW}

The first prescription amounts to a purely additive 
combination,
\beqar
\label{eq:qcdplusew}
\sig{\NNLO\,\QCDpEW}{}
&=&\sig{\LO}{}\left(1+\delt{\QCD}{}+\delt{\EW}{}\right)
+\sig{\LO}{gg}\,,
\eeqar
where all terms of $\ord(\alpha^4)$, $\ord(\alphaS\alpha^4)$, $\ord(\alpha^5)$ and $\ord(\alphaS^2\alpha^4)$
are simply summed. 

\paragraph{\NNLO\,\QCDtEW} 
As a possible approximation of the mixed 
QCD--EW higher-order corrections we consider the 
factorised combination
\beqar
\label{eq:qcdtimesew}
\sig{\NNLO\,\QCDtEW}{}
&=&\sig{\LO}{}\left(1+\delt{\QCD}{}\right)\left(1+\delt{\EW}{}\right)
+\sig{\LO}{gg}\,,
\eeqar
where the EW correction factor is applied to the entire 
NNLO QCD cross section except for the loop-induced $gg$ channel, 
for which the EW corrections $\delt{\EW}{}$ of the
$q\bar q$ and $\gamma\gamma$ channels are not applicable. 
The prescription \refwoeq{eq:qcdtimesew} can also be written
  in the form
\begin{eqnarray}
\label{eq:factpresc2}
\sig{\NNLO\,\QCDtEW}{}
&=&
\sig{\NNLO\,\QCD+\EW}{}+
\sig{\LO}{}
\delt{\QCD}{}\,
\delt{\EW}{}\,
\,.
\end{eqnarray}
Thus, the factorised combination~\refwoeq{eq:factpresc2} generates extra
$\ord(\alphaS\alpha)$ and $\ord(\alphaS^2\alpha)$ mixed QCD--EW corrections. Provided that the dominant sources of QCD and EW corrections 
factorise, such terms can be regarded as a reasonable approximation of
mixed QCD--EW effects.
For instance, at scattering energies 
\mbox{$Q\gg \MW$} this assumption is justified when
EW effects are dominated by Sudakov logarithms, 
and the dominant QCD effects
arise at scales well below $Q$, factorising 
with respect to the underlying \hardVV process.
In such cases, the factorised prescription~\refwoeq{eq:qcdtimesew} should be
regarded as a superior prediction as compared to the additive
combination~\refwoeq{eq:qcdplusew}.

\paragraph{\NNLO\,\QCDtEWqq}

As a motivation for an alternative combination, let us 
highlight the role of individual partonic channels 
in the factorised formula~\refwoeq{eq:qcdtimesew}.
To this end
we rewrite the QCD corrections as
\begin{eqnarray}
\label{eq:QCDsplit}
\sig{\NNLO\,\QCD}{}
&=&
\sig{\LO}{q\bar q}\left(1+\delt{\QCD}{q\bar q}\right)
+\sig{\LO}{\gamma\gamma}+\sig{\LO}{gg}\,,
\end{eqnarray}
where $\delt{\QCD}{q\bar q}$ includes the same QCD corrections as
$\delt{\QCD}{}$, 
but is normalised to the  LO cross section in the $q\bar q$ channel. Moreover
we split the EW corrections into contributions from the $q\bar q$ and $\gamma$-induced
channels,
\begin{eqnarray}
\label{eq:EWsplit}
\sig{\NLO\,\EW}{} &=&
\sig{\LO}{q\bar q}\left(1+\delt{\EW}{q\bar q}\right)
+
\sig{\LO}{\gamma\gamma}\left(1+\delt{\EW}{\gamma\gamma/q\gamma}\right)\,.
\end{eqnarray}
Here in  the factor $\delt{\EW}{q\bar q}$ we include only 
$\ord(\alpha)$ corrections from the $q\bar q$ channel, whereas all other
$\ord(\alpha)$ effects stemming
from the $\gamma\gamma$ and $q\gamma$ channels%
\footnote{This ad-hoc splitting of EW corrections deserves 
some comments.
As pointed out in~\citere{Kallweit:2017khh}, (anti)quark-photon channels 
have the twofold role of EW corrections to the $q\bar q$ and $\gamma\gamma$
channels and are connected to both channels via collinear singularities.
Thus, they cannot be entirely associated with one or the other
channel. 
For this reason,  \refeq{eq:EWsplit} should be understood as 
a purely technical separation of $q\bar q$ and $\gamma$-induced corrections,
which can be adopted upon subtraction of collinear singularities (based on 
dipole subtraction in our implementation).
%
%
As discussed below, the choice of handling the $q \gamma$ channels as 
corrections to the $\gamma\gamma$ channel (rather than to the dominant $q\bar q$
channel) is motivated by the fact that the $q\gamma$ channels can 
lead to giant EW $K$-factors that cannot be combined with the QCD corrections
with a factorised prescription.
\label{footnote:adhoc}
}
are included in the factor $\delt{\EW}{\gamma\gamma/q\gamma}$.
Using the notation of \refeqs{eq:QCDsplit}{eq:EWsplit}
we can rewrite the factorised formula \refwoeq{eq:qcdtimesew} as
\begin{eqnarray}
\label{eq:factsplitting}
\sig{\NNLO\,\QCDtEW}{}
&=&\left[\sig{\LO}{q\bar q}\left(1+\delt{\QCD}{q\bar q}\right)
+\sig{\LO}{\gamma\gamma}\right]\left(1+\delt{\EW}{}\right)
+\sig{\LO}{gg}\,,
\end{eqnarray}
where the EW $K$-factor corresponds to 
\begin{eqnarray}
\delt{\EW}{} &=&
\frac{\delt{\EW}{q\bar q}
\sig{\LO}{q\bar q}
+
\delt{\EW}{\gamma\gamma/\gamma q}
\sig{\LO}{\gamma\gamma}
}{\sig{\LO}{q\bar q}+\sig{\LO}{\gamma\gamma}}\,,
\end{eqnarray}
and can be regarded as the weighted average of the corrections in the $q\bar q$ and
$\gamma\gamma$ channels.
The representation~\refwoeq{eq:factsplitting}
demonstrates that the factorised combination 
does not induce any $\ord(\alphaS)$ effect in the 
$\gamma\gamma$ and $gg$ channels.
The only nontrivial factorised correction
arises from the term $\delt{\QCD}{q\bar q}\delt{\EW}{}$, where
QCD corrections to the $q\bar q$ channel are combined with the average
EW corrections in the $q\bar q$ and $\gamma\gamma$ channels.
The latter includes contributions from 
$q \gamma$ channels that can give rise to giant EW $K$-factors,
in which case a factorised treatment is not justified
(see~\refse{sec:results_inclusive} for a detailed discussion).
For this reason 
we consider the alternative combination formula
\begin{eqnarray}
\label{eq:qcdtimesewqq}
\sig{\NNLO\,\QCDtEWqq}{}
&=&\sig{\LO}{q\bar q}\left(1+\delt{\QCD}{q\bar
q}\right)\left(1+\delt{\EW}{q\bar q}\right)
+\sig{\LO}{\gamma\gamma}
\left(1+\delt{\EW}{\gamma\gamma/q
\gamma}\right)
+\sig{\LO}{gg}\,,
\end{eqnarray}
where the factorisation of EW corrections is restricted to the 
$q\bar q$ channel, while photon-induced channels
and the loop-induced $gg$ contribution are treated in an additive way.
In analogy with \refeq{eq:factpresc2}, 
the prescription \refwoeq{eq:qcdtimesewqq} can be rewritten as\footnote{Note that 
$\sig{\LO}{}\delt{\QCD}{}=
\sig{\LO}{q\bar q}\delt{\QCD}{q\bar q}=
\sig{\NNLO\,\QCD}{}-
\sig{\LO}{}-
\sig{\LO}{gg}$.
See \refeqs{eq:LOdiff}{eq:QCDcorr1} and
(\ref{eq:QCDsplit}).}
\begin{eqnarray}
\label{eq:factpresc2B}
\sig{\NNLO\,\QCDtEWqq}{}
&=&
\sig{\NNLO\,\QCDpEW}{}
+
\sig{\LO}{}
\delt{\QCD}{}\,
\delt{\EW}{q\bar q}\,\,.
\end{eqnarray}

Both multiplicative combinations \refwoeq{eq:factpresc2}
and~\refwoeq{eq:factpresc2B}
are implemented at the level of individual distributions
by computing the relevant differential EW $K$-factors
$\delt{\EW}{}$ and $\delt{\EW}{q\bar q}$ on a bin-by-bin basis.

When QCD corrections are dominated by hard
effects that do not factorise with respect to the \hardVV subprocess, like in the case of 
giant $K$-factors, the difference between the additive and the modified multiplicative
combination can be regarded as a rough indication of the magnitude
of potential effects of $\ord(\alphaS\alpha)$ and beyond.
More details on uncertainty estimates of missing mixed
\QCD--\EW corrections will be discussed in \refse{sec:pheno}.
As far as pure QCD uncertainties are concerned, they are estimated
  through customary variations of the renormalisation and factorisation
scales.
Uncertainties from missing EW corrections beyond $\ord(\alpha)$ are not addressed in this paper:
the dominant source of $\ord(\alpha^2)$ effects at high energy 
are two-loop Sudakov logarithms of the form $\alphaW^2\log^4(Q^2/\MW^2)$, 
which should be included in order to achieve few-percent accuracy 
at high $\pT$. 
The expected size of these two-loop EW effects,
assuming naive Sudakov exponentiation, 
is around $\frac{1}{2}\delta^2_\EW$.

\section{Phenomenological results}
\label{sec:pheno}

In this section we present numerical results for the selected diboson 
processes 
\beqar
\label{eq:dibosonprocs}
pp&\to&\ell^-\ell^+\nu_{\ell'}\bar\nu_{\ell'}\qquad(\zz)\,,\label{proc:zz}\\
pp&\to&\ell^-\ell'^+\nu_{\ell'}\bar\nu_{\ell}\qquad(\ww)\,,\label{proc:ww}\\\
pp&\to&\ell^-\ell^+\ell'^\pm\nu_{\ell'}\qquad(\wz)\,.\label{proc:wz}\
\eeqar
All cross sections correspond to the contribution from one lepton family
 \mbox{$\ell,\ell'=e$} or $\mu$, and \mbox{$\ell'\neq \ell$}.
In the case of \wz production, the QCD and EW corrections are combined at the level of the individual \wpz and \wmz subprocesses, and their cross sections are summed up afterwards.

\subsection{Setup}\label{sec:setup}

In the following we specify the employed input parameters, scale
choices, PDFs, and selection cuts.

\paragraph{Input parameters and schemes}

\begin{table}
  \begin{center}
    \begin{tabular}{rclrcl}
      $\GF$ & \shortequal & $1.1663787\cdot 10^{-5}~\GeV^{-2}$\, & \qquad\qquad & &\\
      $\MW$ & \shortequal & $80.385~\GeV$\,  & $\GW$ & \shortequal & $2.0897~\GeV$\, \\
      $\MZ$ & \shortequal & $91.1876~\GeV$\, & $\GZ$ & \shortequal & $2.4955~\GeV$\, \\
      $\MH$ & \shortequal & $125~\GeV$\,     & $\GH$ & \shortequal & $4.07~\MeV$\, \\
      $\Mb$ & \shortequal & $0\,(4.75)~\GeV$\,    & $\Gb$ & \shortequal & $0$\, \\
      $\Mt$ & \shortequal & $173.2~\GeV$\,   & $\Gt$ & \shortequal & $1.339~\GeV$\,, \\
    \end{tabular}
  \end{center}
  \caption{Values of the various physical input parameters. The value of
  $\Mb$ depends on the employed flavour-number scheme as discussed in the
  text.}
    \label{tab:inputs}
\end{table}

The values of the employed coupling constants, masses and widths are 
listed in \refta{tab:inputs}.
The value of
$\Mb$ depends on the employed flavour-number scheme. 
For \zz and \wz production we use the five-flavour scheme with 
\mbox{$\Mb=0$}, while in the case of \ww production we adopt the four-flavour scheme with 
\mbox{$\Mb=4.75$\,GeV}. This 
renders real-emission channels with bottom quarks in the final state
separately finite, allowing us to remove such channels 
from our predictions. 
In this way, the \ww cross section can be defined without any contamination from resonant
top-quark contributions that would otherwise enter the higher-order QCD corrections,
as discussed in detail in~\citeres{Gehrmann:2014fva,Grazzini:2016ctr}, and also the higher-order EW
corrections, as discussed in~\citere{Kallweit:2017khh}.

Unstable particles are treated in the complex-mass scheme~\cite{Denner:2005fg,Buccioni:2019sur},
where width effects are absorbed into the complex-valued renormalised squared masses
\beq\begin{split}\label{eq:complexmasses}
\mu^2_i\,=&\;\;M_i^2-\ri\Gamma_iM_i
\qquad\mbox{for}\;i=\PW,\PZ,\PH,\Pt\;.
\end{split}\eeq
The complex-mass scheme ensures a gauge-invariant description of non-resonant and off-shell effects in the entire phase space.
The \EW couplings are derived from the gauge-boson masses 
in the \mbox{$\GF$-scheme using}
\beq
\alpha\vert_{\GF} = \frac{\sqrt{2}}{\pi} \, \GF \left| \mu_W^2\,\sin\theta_W^2 \right| \,,
\eeq
where  $\GF$ is the Fermi constant, and 
the complex-valued weak mixing angle is defined through 
\beq
\sin^2\theta_W=1-\frac{\mu_\PW^2}{\mu_\PZ^2}\,.
\eeq
The $G_\mu$-scheme guarantees an optimal description of pure SU(2) interactions at the \EW scale 
by resumming large logarithms of the light-fermion masses 
associated with $\Delta \alpha(M^2_Z)$ and $m_t^2/M_W^2$ enhanced EW
corrections related to $\Delta r$, which summarises the weak radiative corrections to the muon decay~\cite{Sirlin:1980nh}.
This scheme is used for all partonic channels including photon-induced ones.
In this respect we note that, at variance  with on-shell final-state photons,
the appropriate coupling for initial-state photons is not $\alpha(0)$,
but $\alpha$ at a scale of the order of the factorisation scale~\cite{Harland-Lang:2016lhw,Kallweit:2017khh},
as is the case in the $G_\mu$-scheme.\footnote{In general, any EW scheme available within
\OpenLoops~\cite{Buccioni:2019sur} is accessible through the \Matrix{}+\OpenLoops{} framework, 
such that also the $\alpha(M_Z)$-scheme, the $\alpha(0)$-scheme, or mixed $\alpha(0)$--short distance schemes
(relevant for processes with resolved final-state photons)
can be employed.
The various schemes can be selected through the {\tt model.dat} input card
of \Matrix{}, where the default (and recommended) choice is the $\GF$-scheme.  
%
}

The CKM matrix is assumed to be diagonal.
In fact, the negligible mixing of the first two and the third quark 
generations justifies a block-diagonal approximation with Cabibbo-type mixing.
Since all quarks of the first two generations are treated as massless, the unitarity of the CKM matrix ensures the independence 
of all physical results of the Cabibbo angles for charge-neutral processes.\footnote{This statement
  is exact for \zz production at all considered orders, while for \ww production a tiny dependence
  is introduced at NNLO QCD through double-real contributions in the $qq/q\bar q/\bar q\bar q$
  channels through subprocesses of the type of \mbox{$u\bar d\to W^+W^-c\bar s$}.}
Hence, for \zz{} and \ww{} production the choice of a diagonal CKM matrix is fully 
justified.
The effects from a non-trivial CKM matrix for \wz production are also quite small, 
being below $-\percent{1}$ in case of the NLO QCD cross section and
negligible for the NLO QCD $K$-factor \cite{Grazzini:2016swo,Grazzini:2017ckn}. 

\paragraph{Renormalisation and factorisation scales}

In order to avoid large shape corrections in 
the tails of kinematic distributions, we use dynamic QCD scales.
Specifically, the renormalisation scale $\mur$ and the
factorisation scale $\muf$ are set to
\beq\begin{split}\label{eq:RFscales} 
\murf\,=&\;\xi_{\rR,\rF}\,\mu_0\;,
\quad\mbox{with}\quad 
\mu_0\,=\;\tfrac{1}{2}\,\left(E_{\rT,V_1}+E_{\rT,V_2}\right)
\quad\mbox{and}\quad 
\tfrac{1}{2}\le \xi_{\rR},\xi_{\rF}\le 2\;.
\end{split}\eeq 
Here, the transverse energies of the leading and subleading vector bosons,
$V_1, V_2$, are
defined as 
\beq\begin{split}\label{eq:et} 
E_{\rT,V_i}=\sqrt{p_{\rT,V_i}^2+m_{V_i}^2}\,,
\end{split}\eeq
where $p_{\rT,V_i}$ and $m_{V_i}$ are the 
transverse momentum and the invariant mass of the off-shell 
$\ell^+\ell^-/\nu\bar\nu$ pair for a $Z$-boson (or photon in case of $\ell^+\ell^-$) and
a $\ell^+\nu_{\ell}/\ell^-\bar\nu_{\ell}$ pair for a reconstructed
$W^{\pm}$-boson.\footnote{For the different-flavour processes considered in this paper this 
scale choice is unambiguously defined.
For the same-flavour processes, also available in \Matrix, 
corresponding scale choices are implemented via 
reconstruction of the intermediate vector bosons according to a distance criterion.
}
At NLO EW, in order to guarantee infrared safety, the above 
dynamic scale is determined using the momenta of dressed leptons 
as defined below in \refeq{eq:leptondressing}.
Our central scale corresponds to \mbox{$\xi_{\rR}=\xi_{\rF}=1$},
and theoretical uncertainties due to missing higher-order QCD contributions
are estimated through customary 7-point variations \mbox{$(\xi_\rR,\xi_\rF)=(2,2)$},
$(2,1)$, $(1,2)$, $(1,1)$, $(1,\tfrac{1}{2})$, $(\tfrac{1}{2},1)$, 
$(\tfrac{1}{2},\tfrac{1}{2})$.

The multiplicative combinations of QCD and EW corrections
involve EW $K$-factors of the form
\mbox{$\delt{\EW}{}=\sig{\NLO\,\EW}{}(\mu_R,\mu_F)/\sig{\LO}{}(\mu_\rR,\mu_\rF)$},
where scale variations are correlated between the numerator and the denominator. However, 
the $\mu_\rR$ dependence is completely absent, while the 
$\mu_\rF$ dependence cancels almost exactly in the ratio.
As a consequence, the multiplicative combinations of QCD and EW corrections 
results in essentially the same QCD scale uncertainties as for the underlying QCD cross sections. 

\paragraph{Parton densities}
As parton distribution functions (PDFs) for the calculation of hadron-level cross sections we employ, 
consistent with the chosen \mbox{$n_f=5$} or \mbox{$n_f=4$} flavour schemes, the corresponding
NNPDF31\_nnlo\_as\_0118\_luxqed set \cite{Bertone:2017bme}, which
is based on the LUXqed methodology~\cite{Manohar:2016nzj} for the
determination of the photon flux.~\footnote{See~\cite{Harland-Lang:2019eai} for  an alternative treatment of photon-induced processes via direct application of the structure function formalism avoiding parton densities for initial-state photons.}

\paragraph{Selection cuts}
For the infrared-safe definition of 
observables at NLO EW, 
collinear charged-lepton--photon pairs 
within a cone of radius \mbox{$\Rrec = 0.1$}
are systematically recombined 
into dressed leptons according to\footnote{In case this condition holds for more than one lepton, the lepton closest to the photon in $\Delta R_{\ell\Pa}$ is dressed.}
\beq
\label{eq:leptondressing}
p^\mu_{\ell,\mathrm{dressed}} = p^\mu_\ell
+p^\mu_\gamma\qquad\mbox{for}\quad
\Delta R_{\ell\Pa} \,=\, \sqrt{\Delta \phi^2_{\ell\Pa}
+\Delta \eta^2_{\ell\Pa}}  \,<\, \Rrec \,.
\eeq
All cuts, observables and scales are defined in terms of dressed leptons.

We apply uniform fiducial cuts for \zz, \ww, and \wz production
that mimic the standard selections imposed in experimental measurements. 
Specifically, we require
\beq 
\label{cuts:pt}
p_{\rT,\ell^\pm} > 25\,\GeV\,,\quad 
|\eta_{\ell^\pm}| < 2.5 
\eeq
for the transverse momentum and pseudorapidity
of each dressed charged lepton. For processes with neutrinos in the final state we 
additionally require
\beq 
\label{cuts:met}
\missingPT > 25\,\GeV 
\eeq
for the missing transverse momentum calculated as the vectorial sum
of the neutrino momenta. 
Moreover, the invariant mass of same-flavour $\ell^+\ell^-$ pairs 
is restricted to the $Z$-mass window
\beq 
\label{cuts:invm}
66\; \GeV < m_{\ell^+\ell^-} < 116\; \GeV\,.
\eeq

\paragraph{Reconstructed vector bosons}
In the following we present 
differential distributions in 
the transverse momenta and invariant masses of the 
vector bosons.
Such observables are defined in terms of the reconstructed 
vector-boson momenta
\begin{eqnarray}
\label{eq:reco}
p^\mu_Z &=& 
p^\mu_{\ell^+,\mathrm{dressed}}+p^\mu_{\ell^-,\mathrm{dressed}}\quad\mbox{or}\quad
p^\mu_{\nu_\ell}+p^\mu_{\bar\nu_\ell}\,,
\nonumber\\
p^\mu_{W^+} &=& 
p^\mu_{\ell^+,\mathrm{dressed}}+p^\mu_{\nu_\ell}\,,
\nonumber\\
p^\mu_{W^-} &=& 
p^\mu_{\ell^-,\mathrm{dressed}}+p^\mu_{\bar\nu_\ell}\,,
\end{eqnarray}
where all charged leptons are potentially dressed.
Here, the vector bosons are kept off-shell, and the 
correctness of the reconstruction is guaranteed by 
pairing charged leptons and 
neutrinos of the same generation,
 selecting the appropriate neutrino and/or or anti-neutrino 
 momenta at truth level.
The reconstructed vector bosons are ordered according to their 
$\pT$, and the leading and subleading boson are labelled 
$V_1$ and $V_2$, respectively. 

\paragraph{Jet veto}
As discussed in \sct{sec:giantK}, in order to avoid giant $K$-factors at high $\pT$,
we impose selection cuts that single out regions dominated 
by \hardVV production while suppressing regions dominated by \hardVj production.
To this end we apply a jet veto. More precisely, we impose a
restriction on the total jet transverse energy,
\beq
\HTjet\,=\,\sum_{i\in {\rm jets}} p_{\rT,i}\,,
\eeq
where we include all reconstructed anti-$k_{\rm T}$ 
jets~\cite{Cacciari:2008gp} with \mbox{$R=0.4$}, \mbox{$|y|<4.5$}, and arbitrary $\pT$.
The upper bound for $\HTjet$ is defined in terms of the 
hardness of the diboson system. Specifically, we use 
the total leptonic transverse energy,
\beq
\label{cuts:jetveto0}
\HTl \,=\, \sum_{i\in \{\ell^\pm\}} p_{\rT,i} + \missingPT\,,
\eeq
and require 
\beq
\label{cuts:jetveto}
\HTjet  \,<\, \xiveto~\HTl\,,\qquad \mbox{with}\quad
\xiveto=0.2\,.
\eeq
In order to investigate the effect of
giant $K$-factors and their interplay with EW corrections, in 
\refses{sec:results_inclusive}{sec:results_veto}
we will compare
results with and without the above jet veto.
Note that imposing a jet veto on QCD (and QED) radiation may in principle generate 
large logarithms of soft and collinear origin, thereby leading to 
significant uncertainties beyond (N)NLO.
However, the dynamic veto condition
\refwoeq{cuts:jetveto} does not lead to such large
logarithms since 
soft/collinear radiation in the 
region \mbox{$\HTjet/\HTl\ll 1$} is never vetoed.

The effect of the above jet veto on the relative hardness of the 
two vector bosons at high $\pT$ can be quantified by translating 
\refeq{cuts:jetveto} into a lower bound for the $\pT$ of the 
softer vector boson. This is easily achieved by combining 
\refeq{cuts:jetveto} with 
\beqar
\label{cuts:jetveto2}
\big|\vec p_{\rT,V_2}\big|&=& 
\Big| \vec p_{\rT,V_1}+ \sum_{i\in\{q,\bar q, g\}} \vec p_{\rT,i}\Big|
\,\ge \,  p_{\rT,V_1}- \sum_{i\in\{q,\bar q, g\}} p_{\rT,i}
\,\simeq \,  p_{\rT,V_1}-\HTjet\,,
\eeqar
which leads to 
\beqar
\label{cuts:jetveto3}
p_{\rT,V_2}&\ge & p_{\rT,V_1}-\xiveto \HTl\,.
\eeqar
This inequality can be further refined by relating $\HTl$
to the transverse momenta of the two vector bosons. 
To this end we can use
\beqar
\label{cuts:jetveto4}
\HTl= 
\sum_{i\in \{\ell^\pm\}}
 p_{\rT,i}
+
\Big|\sum_{j\in \{\nu,\bar\nu\}} 
\vec p_{\rT,j}\Big|
&\le &
\sum_{i\in \{\ell^\pm,\nu\bar\nu\}}
p_{\rT,i}
\,=\,
H_{\rT,V_1}+H_{\rT,V_2}\,,
\eeqar
where $H_{\rT,V_i}$
denotes the total transverse energy
of the decay products of the $V_i$ vector boson.
In the following we assume that both vector bosons
are nearly on-shell. Moreover, we focus on the region
\beqar
\label{eq:ptmvlimit}
p_{\rT,V_1}\gg M_{V_{1,2}}\,, 
\eeqar
where the products of the 
decay of the leading boson, \mbox{$V_1\to a b$}, are nearly collinear.
Thus
\beqar
\label{cuts:jetveto5a}
H_{\rT, V_1} &=& 
{p}_{\rT,a}
+
{p}_{\rT,b}
\,\simeq\, p_{\rT,V_1}\,.
\eeqar
For the decay of the softer boson, \mbox{$V_2\to cd$}, 
we can use
\beqar
\label{cuts:jetveto5}
H_{\rT, V_2} &=& 
{p}_{\rT,c}
+
{p}_{\rT,d}
\,\le\,
E^\prime_c+E^\prime_d
\,=\, E^\prime_{V_2}
\,=\,  
\sqrt{p_{\rT,V_2}^2+M_{V_2}^2}\,,
\eeqar
where the inequality holds for energies $E^\prime_i$ in any 
reference frame that is
connected to the laboratory frame via a longitudinal boost, while the last
identity is based (without loss of generality) on the reference frame where 
the longitudinal component of $\vec p_{V_2}$ vanishes.
In this way we arrive at 
\beqar
\label{cuts:jetveto6}
\HTl &\le&  
p_{\rT,V_1}+
\sqrt{p_{\rT,V_2}^2+M_{V_2}^2}
\,\simeq\,
p_{\rT,V_1}+p_{\rT,V_2}\,,
\eeqar
using \eqn{eq:ptmvlimit}.
Thus, combining eqs.~\refwoeq{cuts:jetveto3} and \refwoeq{cuts:jetveto6} leads to the
bound 
\beqar
\label{cuts:jetveto7}
{p_{\rT,V_2}}
&\ge & 
\frac{1-\xiveto}{1+\xiveto}
\,{p_{\rT,V_1}}
\,=\,\frac{2}{3}\,
{p_{\rT,V_1}}
\qquad
\mbox{for}\quad \xiveto=0.2\,,
\eeqar
which confirms that the two bosons are similarly hard.
As demonstrated in \refse{sec:results_veto}, 
this bound is violated only by highly suppressed
off-shell contributions. Moreover, at very high transverse momenta, 
the ratio between the $\pT$ of the softer and harder 
vector bosons is typically well above $2/3$ and exceeds $0.9$ on average.

\subsection{Fiducial cross sections}
\label{sec:results_xsections}

\renewcommand\arraystretch{1.4}
\begin{table}
\begin{center}
\input{XStable_baseline.tex}
\end{center}
\caption{Integrated cross sections for the processes \refwoeq{proc:zz}--\refwoeq{proc:wz}
at \mbox{$\sqrt{s}=13$\,TeV}. The baseline cuts~\refwoeq{cuts:pt}--\refwoeq{cuts:invm} are applied 
without jet veto. The quoted uncertainties correspond to seven-point scale
variations.\\
}
\label{tab:baseline_inclusive}
\end{table}

Predictions and scale variations for the fiducial cross sections of the
diboson  processes~\refwoeq{proc:zz}--\refwoeq{proc:wz} are presented in
\refta{tab:baseline_inclusive}.  All results correspond to $pp$ collisions
at \mbox{$\sqrt{s}=13$\,TeV} with the acceptance cuts~\refwoeq{cuts:pt}--\refwoeq{cuts:invm}
and without jet veto.
Results at the various orders in the QCD and EW expansions are shown
separately and combined according to the three different prescriptions
defined in \refse{sec:combination}.  The last three rows of 
\refta{tab:baseline_inclusive} show
the effect of the combinations as relative deviation with respect to NNLO QCD.

The behaviour of QCD and EW corrections in \refta{tab:baseline_inclusive} is
consistent with the well-known results in the literature. The NLO EW
corrections amount to about \percent{-6} for \zz production, and only \percent{-2} and \percent{-3}
for \ww and \wz production, respectively.
The NLO QCD corrections range from
\percent{+36} for \zz production up to \percent{+73} for \wz production.  In
the latter case, the huge NLO effect is due to the presence of an
(approximate) radiation zero at
LO~\cite{Mikaelian:1979nr}.
The NNLO QCD corrections are again positive and vary between 
\percent{+11} and \percent{+16}. The largest NNLO effects
are found in the case of neutral final states, where the contributions from 
loop-induced $gg$ channels are sizeable.

As discussed in the following, the sizeable impact of QCD corrections has
non-negligible implications on their combination with 
EW corrections.
Comparing NNLO QCD
and combined predictions, we observe that the multiplicative prescriptions
yield, as expected, relative correction factors that are very close to the
NLO EW $K$-factor.
Restricting the factorisation to the $q\bar q$ channel tends to increase the
net EW effect by at most one percent in the \wz case.
In contrast, the additive prescription leads to a significant reduction of
the EW effects by up to \percent{2}.  This is due to the fact that
in the additive combination \refwoeq{eq:qcdplusew}
the sizeable higher-order QCD correction $\sig{\LO}{}\delt{\QCD}{}$ remains free from EW corrections.
In this respect, it is reasonable to assume that 
{$\sig{\LO}{}\delt{\QCD}{}$} should receive a comparable EW 
$K$-factor as the LO cross section. For instance, any
short-distance higher-order EW effect 
associated with the SU(2)$\times$U(1)
couplings is expected to factorise with respect to the QCD corrections. 
Thus, using a factorised combination 
appears to be better motivated than just omitting any mixed QCD--EW effect.
Nonetheless, in the absence of more rigorous 
theoretical arguments, the difference between additive and multiplicative
combinations may be taken as a conservative uncertainty estimate at the level of inclusive
cross sections.

\subsection{Differential distributions without veto against QCD radiation}
\label{sec:results_inclusive}

\begin{figure*}[t]
\centering
\includegraphics[width=\textwidth]{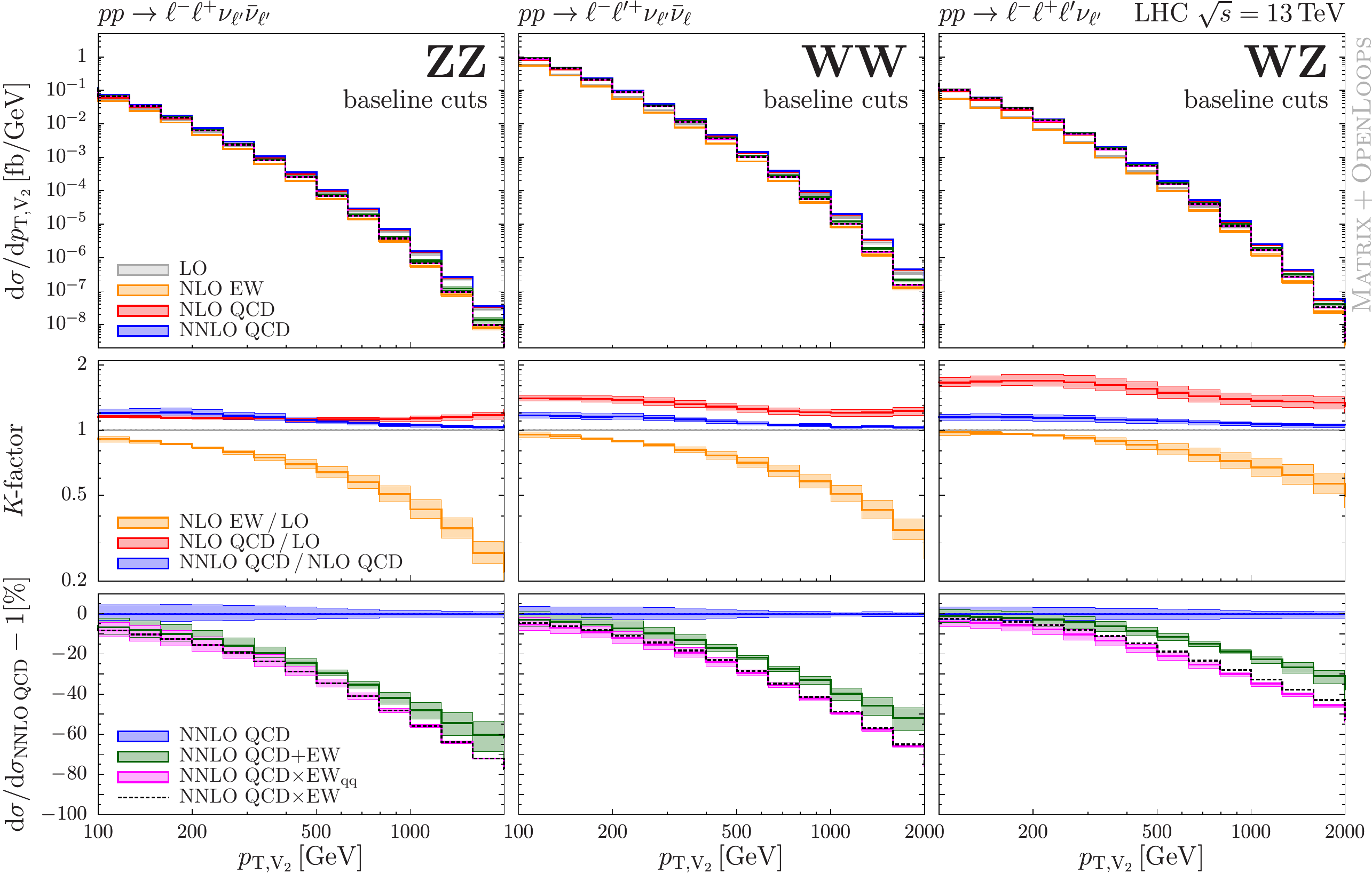}
\caption{Distribution in the transverse momentum of the softer
vector
boson for the \zz
(left), \ww (centre) and \wz (right) production in $pp$ collisions at
13\,TeV.
Baseline cuts without jet veto are applied throughout, and the off-shell vector
bosons are reconstructed according to \refeq{eq:reco}.
The top panel shows absolute predictions at \LO (grey), \NLO \EW (yellow), \NLO \QCD (red)
and \NNLO \QCD (blue).
The centre panel shows NLO EW and NLO QCD 
predictions normalised to LO,  and NNLO QCD predictions normalised to NLO QCD.  
The lower panel shows
the different combinations of \NNLO~\QCD with \NLO~\EW corrections: the \NNLO
\QCDpEW combination~\refwoeq{eq:qcdplusew} (green), the \NNLO \QCDtEW
combination~\refwoeq{eq:qcdtimesew} (dashed black) and the modified \NNLO
\QCDtEW combination~\refwoeq{eq:qcdtimesewqq} (magenta), all normalised to
\NNLO \QCD.  Uncertainty bands correspond to seven-point factor-two 
scale variations as defined
in \sct{sec:setup}, and in the ratios the variation is only applied to the
numerator. Scale variation bands on top of \NNLO~\QCDtEW are omitted in order to
improve legibility. They are basically identical with scale variations at \NNLO \QCD.
}
\label{fig:inc_pTV2} 
\end{figure*}

In \reffis{fig:inc_pTV2}{fig:inc_MET} we present 
differential distributions that characterise the behaviour of the 
vector bosons and their decay products in \zz, \ww and \wz production.
The inclusive cuts~\refwoeq{cuts:pt}--\refwoeq{cuts:invm} are applied throughout
without jet veto, and the vector bosons are reconstructed as detailed in 
\refeq{eq:reco}.
In order to highlight EW Sudakov effects, transverse-momentum and
invariant-mass distributions are plotted up to 
the multi-TeV range,
where event rates are suppressed by several orders of magnitude.  
In this respect, we note that the relevance of our results goes beyond
leptonic diboson signatures at the LHC, where the energy reach at high
luminosity is around 1\,TeV in transverse momenta and 2\,TeV in invariant masses.
In fact,
in the case of hadronic vector-boson 
decays, where statistics is significantly higher,
the QCD and EW corrections are expected
to behave in a rather similar way as for the leptonic channels
discussed in this paper. Indeed, the dynamics of the dominant
giant QCD K-factors and large EW Sudakov corrections only depends on the 
diboson production processes $pp\to VV$ and $pp\to VVj$,
i.e.~it is independent of the vector-boson decays.
Moreover, since EW Sudakov logarithms are mainly
sensitive to the final-state transverse momenta and depend only weakly on
the collider energy, 
the presented results can be regarded as an indication of the
behaviour of EW corrections in the multi-TeV regime 
at future $pp$ colliders.

\begin{figure*}[t] \centering
\includegraphics[width=\textwidth]{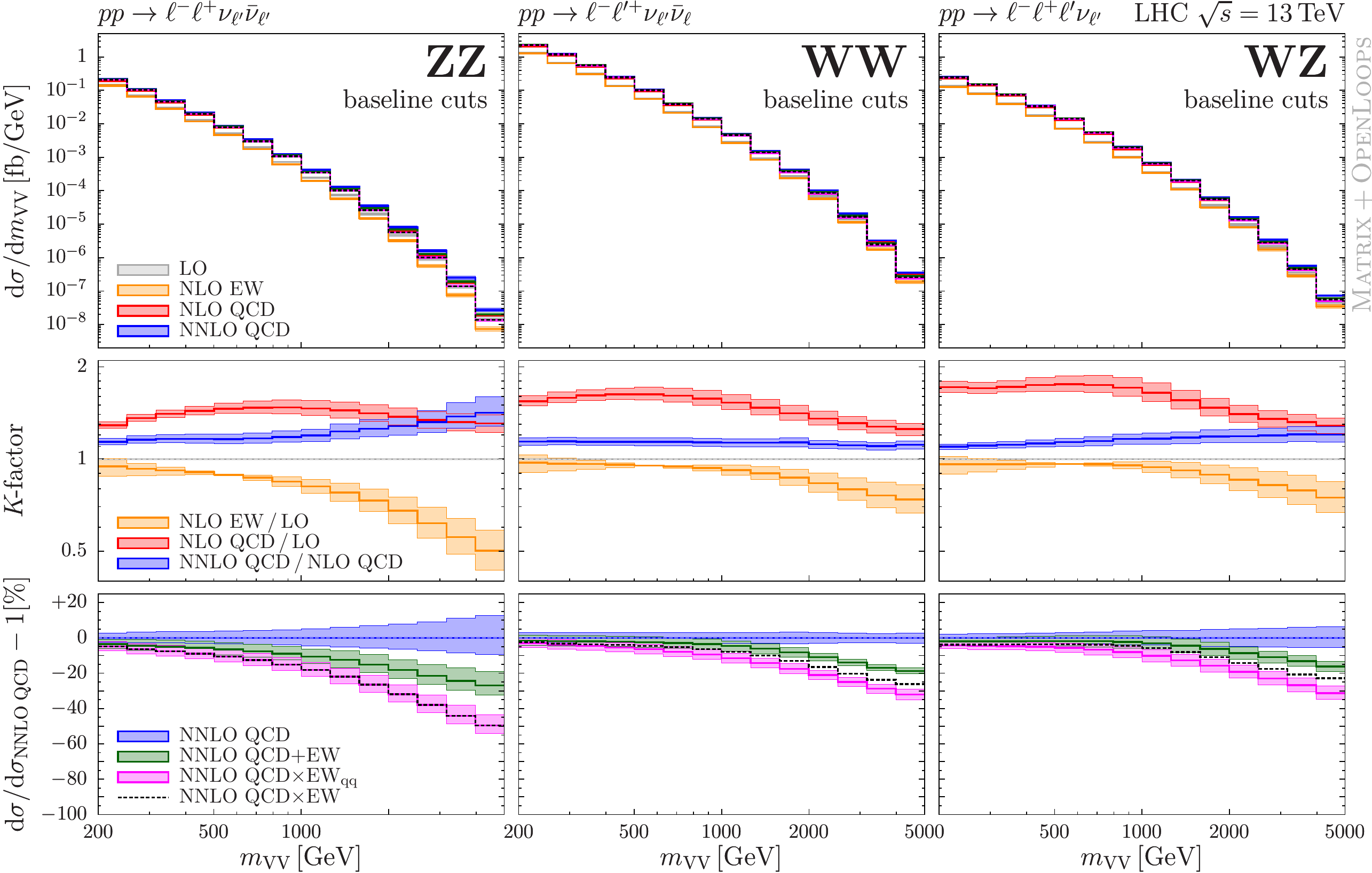}
\caption{ Distribution in the invariant mass of the reconstructed diboson
system for the processes \refwoeq{proc:zz}--\refwoeq{proc:wz} at 13\,TeV. 
Baseline cuts are applied without jet veto.  Plot format and predictions as in \fig{fig:inc_pTV2}.  \label{fig:inc_mVV} } \end{figure*}

The distribution in the $\pT$ of the softer
vector boson  is shown in \reffi{fig:inc_pTV2}.
This observable is an optimal probe of \hardVV production
since requiring large $p_{\rT, V_2}$ automatically ensures 
that also $p_{\rT, V_1}$ is large, thereby excluding giant $K$-factor effects 
due to \hardVj production.
As far as the QCD corrections are concerned, 
at low $p_{\rT, V_2}$  we find similar NLO and NNLO QCD effects 
as in the case of the fiducial cross section, \ie the NLO corrections 
are rather large for \wz production and significantly less pronounced 
for \ww and \zz production, while NNLO effects are at the
\percentrange{10}{20} level for all three processes.
At large $\pT$ higher-order QCD effects tend to decrease.
In particular, the NNLO QCD corrections behave in a very
similar way for all three processes and go down to the few-percent level in 
the TeV range. Also scale uncertainties decrease at large $\pT$, 
being only a few percent in the high-$p_{\rT, V_2}$ tail.
The EW corrections are negative, and in the tails they grow like
$\log^2({p^2_{\rT,V_2}}/\MW^2)$,
as expected for EW Sudakov logarithms in hard $VV$ production.
At 1--2\,TeV the (negative) EW corrections to \zz, \ww and \wz production 
range, respectively,  between \percentrange{60}{75}, \percentrange{50}{65} 
and \percentrange{35}{45}.
Electroweak Sudakov effects remain very large also in the combination of QCD and 
EW corrections, and due to the moderate size of QCD corrections in the tails,  
their impact depends only weakly on the employed combination prescription. 
The two multiplicative combinations are very close to each other, and they are
both well motivated since the observable at hand is dominated by \hardVV
topologies.  The difference between additive and multiplicative combinations
can be regarded as upper bound for the uncertainty due to missing
$\ord(\alphaS\alpha)$ effects.
In the $\pT$ range where NLO EW corrections approach the 
\percent{50} level, the main source of theoretical uncertainty is given by 
missing EW Sudakov logarithms at relative $\ord(\alpha^2)$. Based on naive 
Sudakov exponentiation, their size is expected to be
around $\frac{1}{2}\delta_\EW^2$.

\begin{figure*}[t]
\centering
\includegraphics[width=\textwidth]{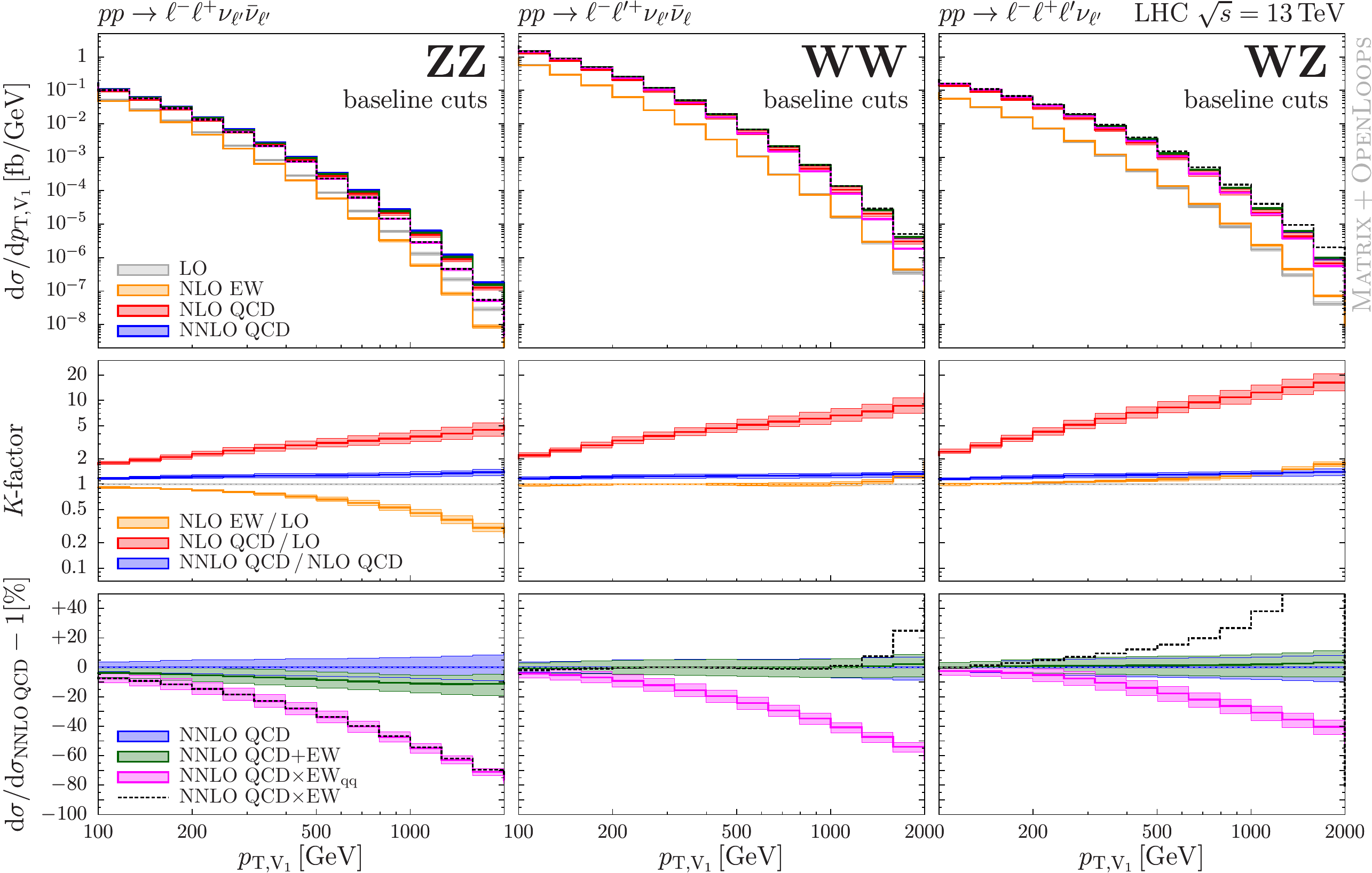}
\caption{
Distribution in the transverse momentum of the harder reconstructed vector boson
for the processes \refwoeq{proc:zz}--\refwoeq{proc:wz} at 13\,TeV. 
Baseline cuts are applied without jet veto.
Plot format and predictions as in \fig{fig:inc_pTV2}.
\label{fig:inc_pTV1}
}
\end{figure*}

In \reffi{fig:inc_mVV} we present the distribution in the 
diboson invariant mass, $m_{VV}$. Again, at small invariant mass we observe similar
QCD corrections as for the fiducial cross sections.
At very large $m_{VV}$ the NLO QCD corrections tend to decrease.
As for the NNLO corrections, in \wz production they increase mildly towards large $m_{VV}$
and reach up to \percent{20} in the TeV regime. In \ww production the additional $gg$ channel, whose contribution decreases at large $m_{VV}$, results in a relatively flat total NNLO correction at around \percent{10}, with residual scale
uncertainties at the \percent{5} level. 
On the other hand, for \zz production we observe a 
significant enhancement of the NNLO corrections up to \percent{40} at very high invariant masses in the multi-TeV range.
This is due to the effect of acceptance cuts and their 
interplay with the opening of the 
\mbox{$qq'\to VV qq'$}, \mbox{$\bar q \bar q'\to VV \bar q\bar q'$} and \mbox{$q\bar q'\to VV q\bar q'$} channels at NNLO.
The reason is that, at large $m_{VV}$, in these channels 
the gauge bosons are mainly emitted in the forward and backward
directions. Hence, the rapidity of their leptonic decay products is often too
large to pass the acceptance cuts on charged leptons in \ww and \wz
production. In contrast, in the case of \zz production the 
absence of rapidity cuts on the \mbox{$Z\to \nu\bar \nu$} decay products
leads to significantly larger contributions from these
channels.
In fact we have checked that the NNLO QCD corrections to \zz,
\ww and \wz production behave in a very similar way when applying the same (technical) cuts on
neutrinos and charged leptons, and switching off loop-induced $gg$ contributions.
 
The EW corrections to the $m_{VV}$ distribution are negative, and in the tails they grow 
like double Sudakov logarithms. However, their impact is less pronounced than for the $p_{\rT,V_2}$ distribution in \reffi{fig:inc_pTV2}. This is due to the fact that 
diboson production at large $m_{VV}$ is dominated by $t$- and $u$-channel 
topologies where the gauge bosons are mainly emitted in the 
forward/backward regions, and the scales $t,u$ that enter Sudakov logarithms
are well below $m_{VV}$.
The largest EW corrections are found in the \zz channel, 
where they amount to \percent{-15} at 
1\,TeV.
In the combination of QCD and EW corrections 
the difference 
between additive and multiplicative prescriptions is similarly large as 
NNLO QCD scale uncertainties, and depending on the process
it can reach up to \percentrange{10}{20} in the multi-TeV region.
For \ww and \wz production we also find a difference of up to
$\percent{5}$ between the two factorised prescriptions.
This effect can be attributed to photon-induced 
\mbox{$\gamma q \to W V q$} channels, where the topologies with 
$t$-channel $W$ bosons that couple to the initial-state photons
(see e.g.~\fig{fig:EWdiagramsWZ}\,l)
yield a significant (positive) NLO EW contribution.

The distribution in the transverse momentum of the harder
vector boson, presented in \reffi{fig:inc_pTV1}, 
shows a completely different behaviour of 
the higher-order effects.\footnote{Fiducial cross sections
that quantify the corrections observed in the tails 
of \reffi{fig:inc_pTV1} are listed in appendix~\ref{app:xsections}.}
At 100\,GeV, the NLO QCD corrections are as large as a factor 
two, and their size grows with $p_{\rT,V_1}$ reaching 
five to twenty times the LO cross section at 2\,TeV.
These giant NLO $K$-factors are driven by \hardVj subprocesses, where 
the recoil of the harder vector boson is absorbed by the hard jet, 
while the second boson becomes soft and generates large logarithmic 
corrections of the form of \eqn{eq:giantkfactA}.
As discussed in \sct{sec:giantK}, such large logarithms 
occur only at NLO and do not need to be resummed to all orders.
With other words, the perturbative expansion beyond NLO is expected to be well
behaved.
This is confirmed by the NNLO QCD results in \reffi{fig:inc_pTV1}, 
where we observe NNLO QCD corrections at the level of \percentrange{30}{50}, which are moderate compared to the giant NLO QCD corrections.
At the same time we find a reduction of scale uncertainties down to the 
\percent{10} level when moving 
from NLO to NNLO QCD. 
In practice, at large $p_{\rT,V_1}$, the \mbox{$pp\to VV$} cross section at NNLO
QCD is largely dominated by \mbox{$pp\to VVj$} at NLO QCD, which provides the first
reliable QCD prediction for the tail of the distribution at hand.

The behaviour of the EW corrections in \reffi{fig:inc_pTV1} depends strongly
on the diboson final state. For \zz production we observe large negative corrections that grow with $\pT$, 
reaching about \percent{-50} with respect to LO at 1\,\TeV.
This indicates that the NLO EW corrections  
are dominated by Sudakov logarithms, which originate 
from the virtual corrections to hard-\zz production.\footnote{Note that in the 
virtual $\ord(\alpha)$ corrections
the two gauge boson are 
equally hard due to the absence of real radiation.} 
However, the tail of the distribution at hand
is dominated by higher-order QCD contributions corresponding to hard-$Zj$ production. As a result, 
when NLO EW effects  
are combined with NNLO QCD predictions using the additive prescription,
their relative weight goes down by up to a factor of seven.
This is due to the fact that the dominant \mbox{$pp\to VVj$} process does not
receive any EW corrections in the additive combination.
In contrast, in the multiplicative combinations the $\delta_\EW$ correction
factor is applied also to \mbox{$pp\to VVj$}, resulting in large EW Sudakov effects.  

In the case of \ww and \wz production, the behaviour of the EW corrections 
at large $p_{\rT,V_1}$ is 
completely different: instead of negative Sudakov effects, we 
observe positive NLO EW $K$-factors that grow with $p_{\rT, V_1}$
and approach a factor two in the case of \wz production.
This is not due to the absence of EW Sudakov logarithms, but to the fact
that they are overcompensated by a positive giant EW $K$-factor stemming from
the emission of one additional (anti)quark at NLO EW.
More precisely, this large positive contributions 
originate from photon-induced \mbox{$\gamma q\to WV q$} 
and \mbox{$\gamma \bar q\to WV \bar q$} 
topologies
with $t$-channel $W$-boson exchange 
(see~\fig{fig:EWdiagramsWZ}\,l)
that we already discussed 
in the context of the $m_{VV}$ distribution.
Such topologies contribute
only to \ww and \wz production and, as a result of the
$WW\gamma$ triple gauge coupling,
they are strongly enhanced as compared to 
\mbox{$\gamma q\to VV q$} topologies
that involve $t$-channel (anti)quark exchange 
(see e.g.~\fig{fig:EWdiagramsWZ}\,k)
and contribute also to \zz production \cite{Baglio:2013toa}. 
Due to the interplay between large positive QCD effects, large 
negative Sudakov EW effects and large positive photon-induced EW effects, 
the combination of QCD and EW corrections
is strongly dependent on the employed prescription. In the additive 
combination the relative impact of EW effects is minimal, due 
to both the large cancellation between positive and negative EW effects 
and the absence of EW corrections to the dominant \hardVj subprocess.
This results in a dramatic underestimate of higher-order EW
contributions.
In contrast, in the factorised combinations 
the EW $K$-factor is applied to the 
full NNLO QCD cross section, 
resulting in sizeable net EW effects in the tails.
In the fully factorised 
prescription the overall effect is positive due to the 
dominance of the $\gamma$-induced bremsstrahlung contributions to the EW corrections. 
However, this prescription is not well justified in the case at hand, since
the large $\gamma$-induced  EW effect is due to the opening of the 
$VVq$ three-body phase space, while 
in the process \mbox{$pp\to VVj$},
which dominates the (N)NLO QCD corrections,
the three-body phase space is already open.
This justifies the restriction of the 
factorisation to the $q\bar q$ channel,
which leads to an overall large negative EW correction 
of Sudakov-type in the tail. Instead, the fully factorised \NNLO \QCDtEW combination should be discarded.

In general, 
 in the $p_{\rT,V_1}$ distribution, the modified multiplicative
prescription \refwoeq{eq:qcdtimesewqq} yields
a Sudakov-type behaviour for all processes. However, in this approach the NLO EW $K$-factor for \hardVV production is effectively transferred to phase space regions that are dominated by \hardVj
production.
This can largely overestimate the EW corrections since, based on the general
properties of Sudakov logarithms,
the EW corrections to \hardVj production are expected 
to be roughly half as large as for \hardVV production.
For this reason, it is reasonable to use as nominal prediction
 the average of the additive combination \refwoeq{eq:qcdplusew}
and the modified multiplicative
combination \refwoeq{eq:qcdtimesewqq}, while
their spread can be interpreted
as $\ord(\alphaS\alpha)$ uncertainty band.

\begin{figure*}[t]
\centering
\includegraphics[width=\textwidth]{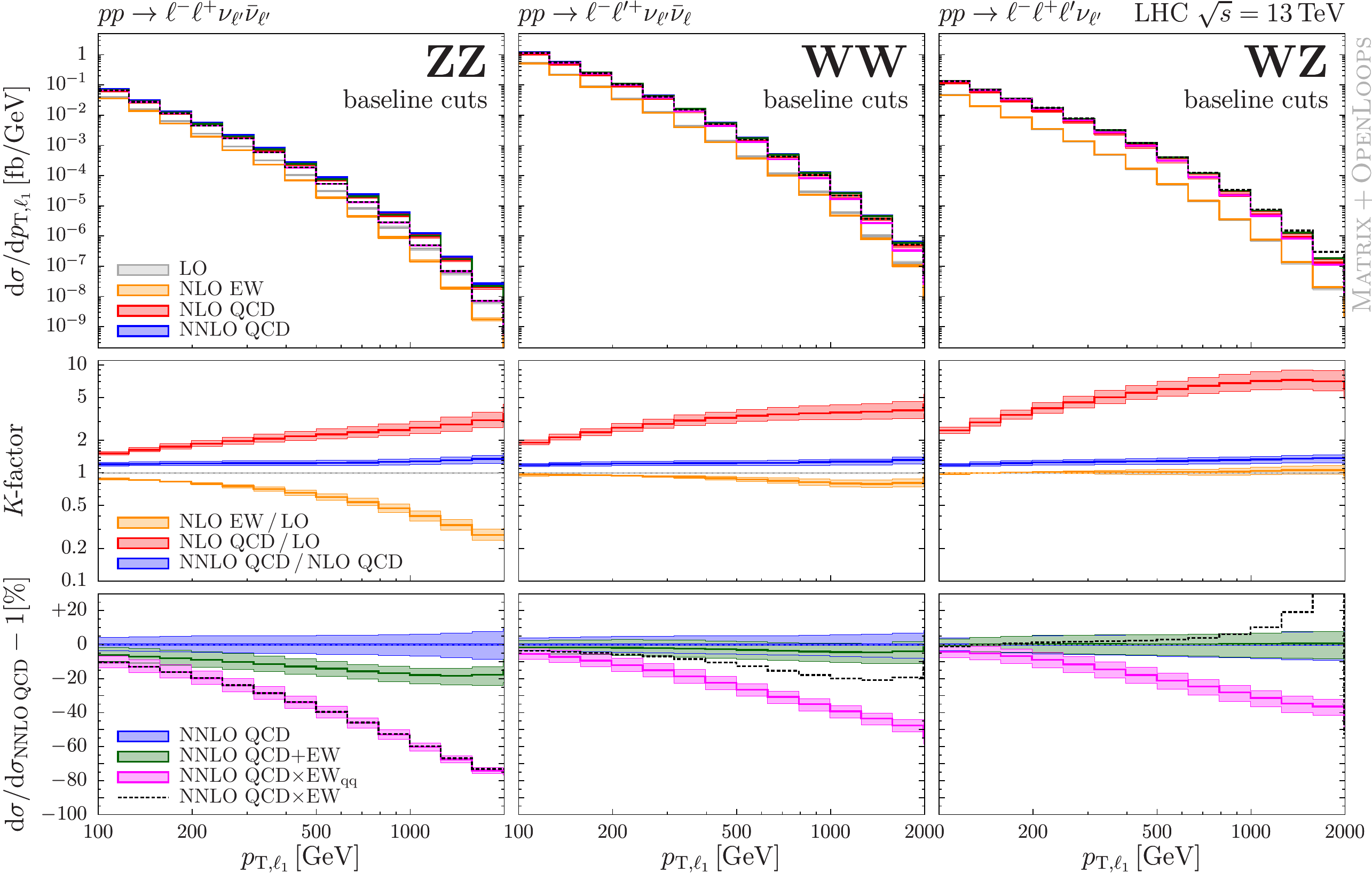}
\caption{
Distribution in the transverse momentum of the hardest charged lepton
for the processes \refwoeq{proc:zz}--\refwoeq{proc:wz} at 13\,TeV. 
Baseline cuts are applied without jet veto.
Plot format and predictions as in \fig{fig:inc_pTV2}.
\label{fig:inc_pTl1}
}
\end{figure*}

\begin{figure*}[t]
\centering
\includegraphics[width=\textwidth]{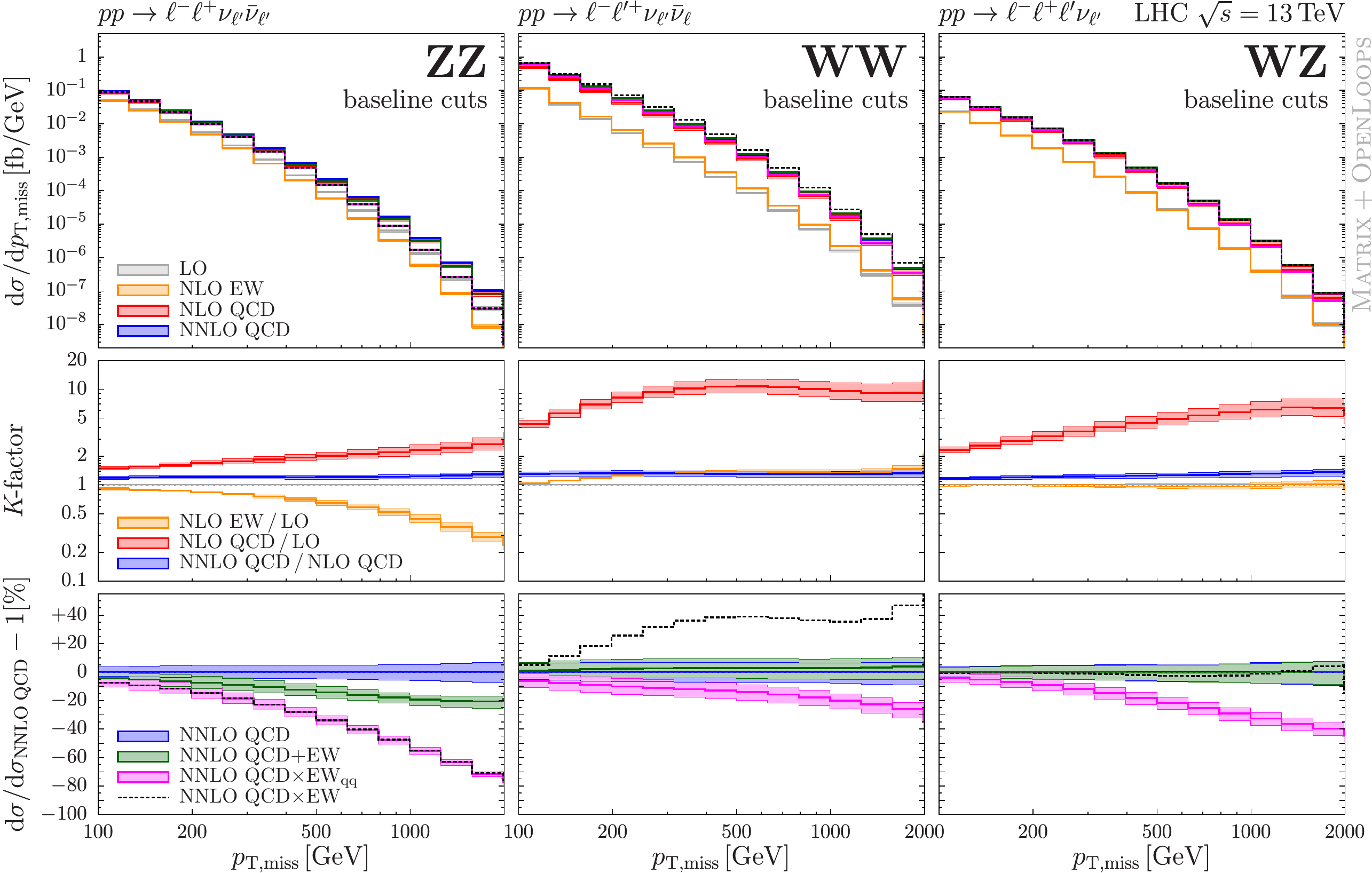}
\caption{
Distribution in the missing transverse momentum
(calculated from the vectorial sum of the neutrino momenta)
for the processes \refwoeq{proc:zz}--\refwoeq{proc:wz} at 13\,TeV. 
Baseline cuts are applied without jet veto.
Plot format and predictions as in \fig{fig:inc_pTV2}.
\label{fig:inc_MET}
}
\end{figure*}

The distribution in the $\pT$ of the
hardest charged lepton, shown in \reffi{fig:inc_pTl1}, features a 
qualitatively similar behaviour of the QCD and EW corrections as for $p_{\rT,V_1}$, shown in
\reffi{fig:inc_pTV1}.
This demonstrates that the issues related to giant $K$-factors 
are independent of the reconstruction of the vector bosons.  
This is substantiated by the distribution in the missing transverse
momentum, $\missingPT$, in \reffi{fig:inc_MET}.
In the \zz and \wz channels, where $\missingPT$
is identical or strongly correlated with the $\pT$ of one of the vector
bosons, the QCD and EW corrections feature qualitatively similar giant
$K$-factor effects as for the distribution in $p_{\rT,V_1}$.
Also in the \ww channel we observe such a behaviour. However, in this case 
the giant $K$-factor effects are already visible at 100\,GeV, where the 
NLO QCD corrections grow very quickly, reaching $\ord(10)$
at few hundred GeV.
As discussed in detail in \citere{Kallweit:2017khh},
this is due to the fact that 
for on-shell \ww production at LO
the vanishing transverse momentum of the diboson system
implies  that \mbox{$\missingPT\le \MW$}.
Hence, at LO the production of a neutrino pair with 
\mbox{$p_{\rT,\nu\bar \nu}=\missingPT \gg  M_W$} emerges
only from strongly suppressed configurations with off-shell vector bosons.
In contrast, thanks to the finite $p_{\rT,WW}$  generated by the 
radiative process \mbox{$pp\to \ww j$}, at NLO QCD
the region \mbox{$\missingPT>\MW$} can be populated with on-shell gauge bosons,
resulting in a much larger cross section with respect to LO.
Analogous giant $K$-factor effects show up also at NLO EW 
through the \mbox{$\gamma q \to WW q$} channel, which leads to a positive 
correction that overcompensates EW Sudakov logarithms and leads 
to an enhancement of a few tens of percent in the tail.
Since such QCD and EW giant $K$-factors are due to the opening 
of the \mbox{$p_{\rT,WW}>0$} phase space, higher-order 
effects beyond NLO are expected to be free from this kind of enhancements.
Indeed this is confirmed by the moderate size  of the NNLO QCD corrections
over the entire $\missingPT$ range.
For what concerns the combination of QCD and EW corrections, similar
considerations as for \reffi{fig:inc_pTV1} apply.

In summary, for observables that are dominated by \hardVV production, such as
the distributions in $p_{\rT,V_2}$ and $m_{VV}$, the factorised combinations
of EW and QCD corrections are well motivated, and $\ord(\alphaS\alpha)$ 
uncertainties are expected to be significantly smaller than 
the differences 
between the factorised and additive prescriptions.
Conversely, in the tails of the distributions that are dominated 
by \hardVj production,\footnote{{We remind the reader that, as
introduced in \sct{sec:giantK}, by \hardVj production we refer to} $pp\to VVj$
with the second vector boson being soft.}
such as for the distributions in $p_{\rT,V_1}$, $p_{\rT,\ell_1}$ and  $\missingPT$,
due to the presence of giant $K$-factors all 
combination prescriptions considered in this paper are inadequate.
In particular, the fully factorised prediction \refwoeq{eq:qcdtimesew} should be
discarded, 
and the band defined by the difference between the additive combination and
the modified multiplicative combination can be considered as
uncertainty related to $\ord(\alphaS\alpha)$ corrections 
and higher, and its average as  nominal prediction.
In the following we show how giant $K$-factors and related issues can be
avoided by means of appropriate selection cuts, while
a more accurate QCD--EW combination in the presence of 
giant $K$-factors is deferred to future work.

\subsection{Results with a veto against hard QCD radiation}
\label{sec:results_veto}

As anticipated in \refse{sec:giantK}, giant $K$-factors can be avoided by
restricting the phase space of \mbox{$pp\to VV$+jets} radiative processes 
to regions dominated by \hardVV production while rejecting
regions dominated by \hardVj production in association 
with soft vector-boson radiation.
This can be achieved either
via direct selection cuts that require similar hardness for the two
vector bosons or through a veto on QCD radiation. In the
following we adopt the latter approach
by  imposing an upper bound on the total jet transverse energy
$\HTjet$, as defined in \refeq{cuts:jetveto}.  

\begin{figure*}[t]
\centering
\includegraphics[width=\textwidth]{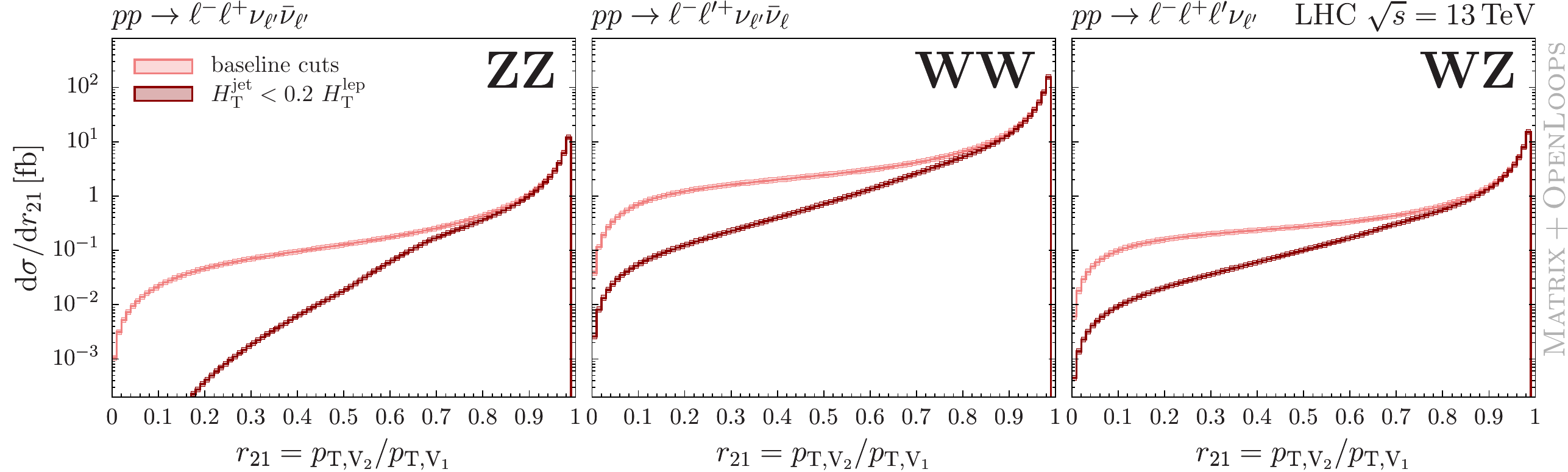}\\[2ex]
\includegraphics[width=\textwidth]{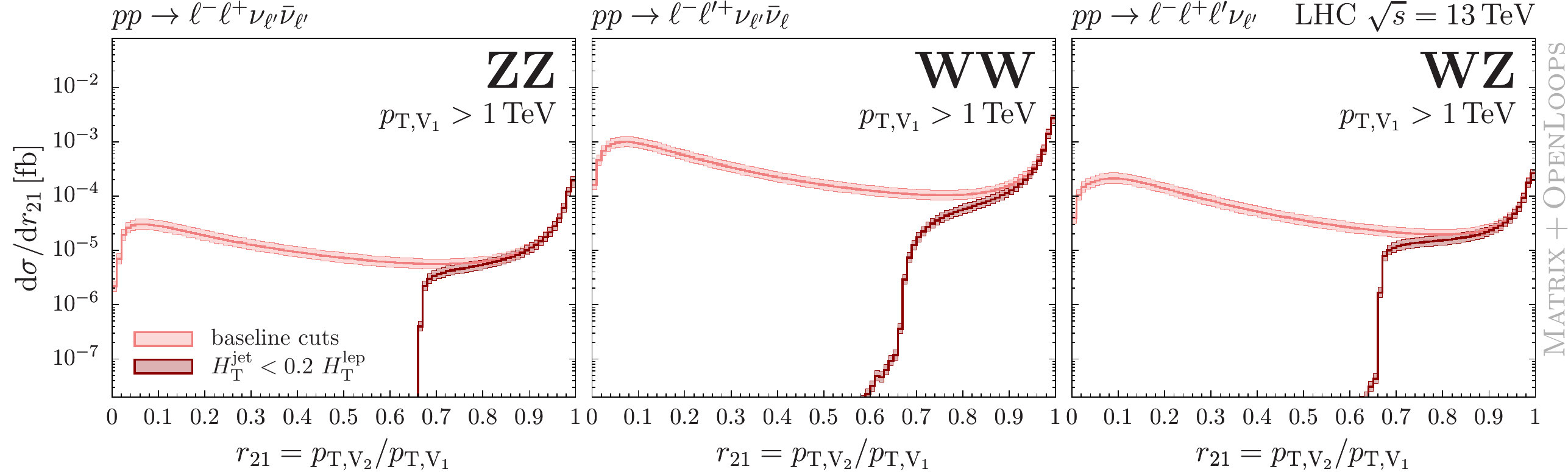}\\
\caption{
  Distribution in the ratio of the transverse momenta of the softer and the harder reconstructed vector bosons,
  \mbox{$r_{21}=p_{\mathrm{T,V_2}}/p_{\mathrm{T,V_1}}$}, at NLO QCD for the processes \refwoeq{proc:zz}--\refwoeq{proc:wz} at 13\,TeV. 
Baseline cuts are applied without (light red) and with (dark red) a jet veto.
In the lower frames, an additional cut on the transverse momentum of the harder vector boson, \mbox{$p_{\mathrm{T,V_1}}>1\,\mathrm{TeV}$} is applied.
\label{fig:r21}
}
\end{figure*}

The interplay of \hardVV and \hardVj production and 
the impact of the jet veto are illustrated in \reffi{fig:r21}, which 
shows NLO QCD predictions for the ratio of the $\pT$ of the softer over the 
one of the harder vector
boson, \mbox{$\rsh=p_{\rT,V_2}/p_{\rT,V_1}$}. 
Due to momentum conservation, at LO   
the two vector bosons are back-to-back, \ie \mbox{$\rsh=1$}, and
the phase space region with \mbox{$\rsh<1$} is populated only through QCD (or QED) radiation starting at NLO.
If only baseline cuts are applied (light red curve in the upper frames), 
the cross section is strongly peaked in the \mbox{$\rsh\to 1$} region,
which corresponds to $VVj$ configurations with 
a \hardVV system accompanied by soft and collinear QCD radiation.
The bulk of the cross section originates from
\mbox{$\rsh>0.8$} and is largely unaffected by the jet veto (dark red curve in the upper frames).
In the hard region \mbox{$p_{\rT,V_1}>1\,\TeV$} (see lower frames)
the shape of the distribution in $\rsh$ is very different.
In the absence of a jet veto,
the cross section develops a very strong enhancement 
in the region of small $\rsh$,
which corresponds to \hardVj production in association with a 
soft vector boson, and is responsible for the
giant $K$-factors discussed in \refse{sec:results_inclusive}.
Note that this \hardVj enhancement exceeds the large-\rsh enhancement 
due to soft QCD radiation in a very broad region, 
from very small $\rsh$ up to \mbox{$\rsh\simeq 0.8$}.
For \mbox{$p_{\rT, V_1}>1\,\TeV$}, we observe that,
consistently with the bound \eqref{cuts:jetveto7}, 
the jet veto acts as a rather sharp cut-off 
rejecting \hardVj configurations with \mbox{$\rsh < 2/3$}.
In fact, the region below $2/3$ is filled 
only by strongly suppressed off-shell contributions.
As a result of the jet veto, the cross section is dominated by the
region of large $\rsh$. Actually, the average value of $\rsh$ is well above 2/3 and exceeds 0.9,
confirming that \hardVV production dominates.
At large $\rsh$ the distribution becomes logarithmically divergent,
and the singularity is cancelled by the virtual contribution at \mbox{$\rsh=1$}. 
Any tight cut in the vicinity of \mbox{$\rsh=1$} would 
induce large logarithmic corrections of soft and collinear origin,
requiring an all-order resummation.
However, the effective $\rsh$-cut imposed through the
jet veto is far enough from the 
logarithmically enhanced region of the
$\rsh$-distribution such that fixed-order predictions are 
well behaved.

\begin{figure*}[t]
\centering
\includegraphics[width=\textwidth]{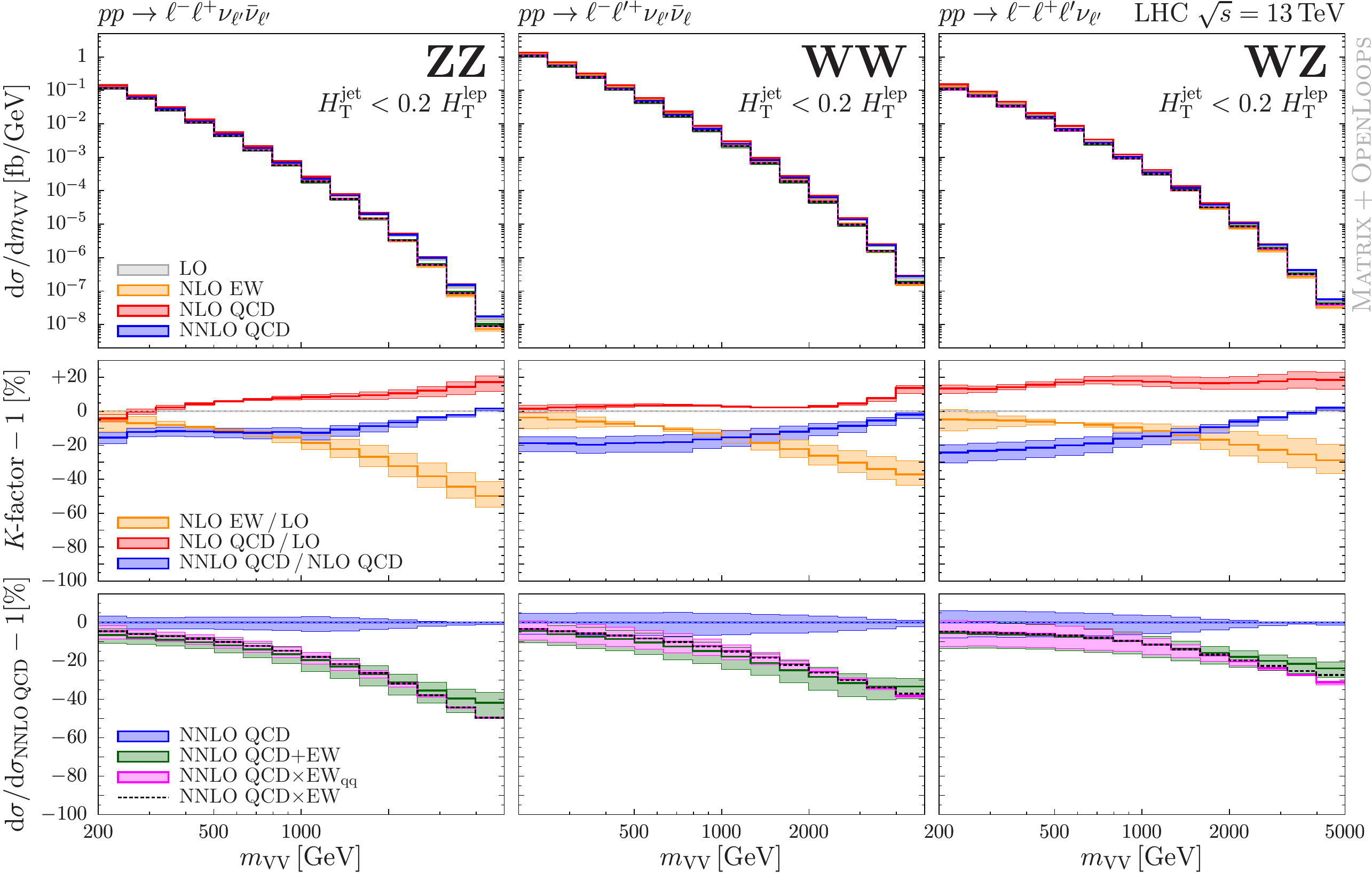}
\caption{
Distribution in the invariant mass of the reconstructed vector-boson pair
for the processes \refwoeq{proc:zz}--\refwoeq{proc:wz} at 13\,TeV. 
Plot format and predictions as in \fig{fig:inc_mVV}, 
but in addition to the baseline cuts also the jet veto~\refwoeq{cuts:jetveto}
is applied.
\label{fig:veto_mVV}
}
\end{figure*}

In \reffis{fig:veto_mVV}{fig:veto_MET} we study the effect of the 
jet veto on the distributions discussed
in \refse{sec:results_inclusive}. 
The veto has no effect on LO predictions, since
it acts only on (N)NLO radiation.
For observables that are free from giant $K$-factors,
the qualitative behaviour of QCD and EW corrections remains 
more or less unchanged, 
apart from a significant reduction of the (N)NLO cross section.
This is illustrated in \reffi{fig:veto_mVV} for the distribution in
$m_{VV}$. At the lower end of the plotted $m_{VV}$ range the 
veto reduces the NNLO QCD cross section by about \percent{50} for all
diboson processes, while its impact is less pronounced at large
$m_{VV}$ and amounts to a reduction of \percentrange{20}{35} in the multi-TeV
region.%
\footnote{These quantitative statements can directly be inferred from the ratio plots in \reffi{fig:inc_mVV} and \ref{fig:veto_mVV}. For example, at low $m_{VV}$ in the case of \zz{} production without jet veto (\reffi{fig:inc_mVV}) the
NLO corrections amount to about \percent{+30} while the NNLO corrections read \percent{+15}, thereby leading to a positive \percent{50} effect. When the jet veto is applied (\reffi{fig:veto_mVV}) the NLO and NNLO effects amount to \percent{-5} and \percent{-15}, respectively. Since the jet veto does not affect the LO result, this amounts to the quoted \percent{50} reduction at low $m_{VV}$.}
This demonstrates 
that the entire distribution at hand is indeed dominated
by \hardVV production also in the inclusive case. 
For this reason, the effect of the jet veto on EW corrections is quite
small: in the case of 
\zz production it is negligible, while for \ww and \wz production it 
enhances the negative correction in the tail by about \percent{5}
due to the 
suppression of the positive \mbox{$\gamma q\to WV q$} contribution
(see the discussion of \reffi{fig:inc_mVV}).
Since NNLO QCD corrections are largest (up to \percent{-25}) for \mbox{$m_{VV}\lesssim 1$\,\TeV}, where EW effects are smaller, while NNLO QCD corrections decrease in the tails, their combination turns out to depend only weakly on the employed prescription.

\begin{figure*}[t]
\centering
\includegraphics[width=\textwidth]{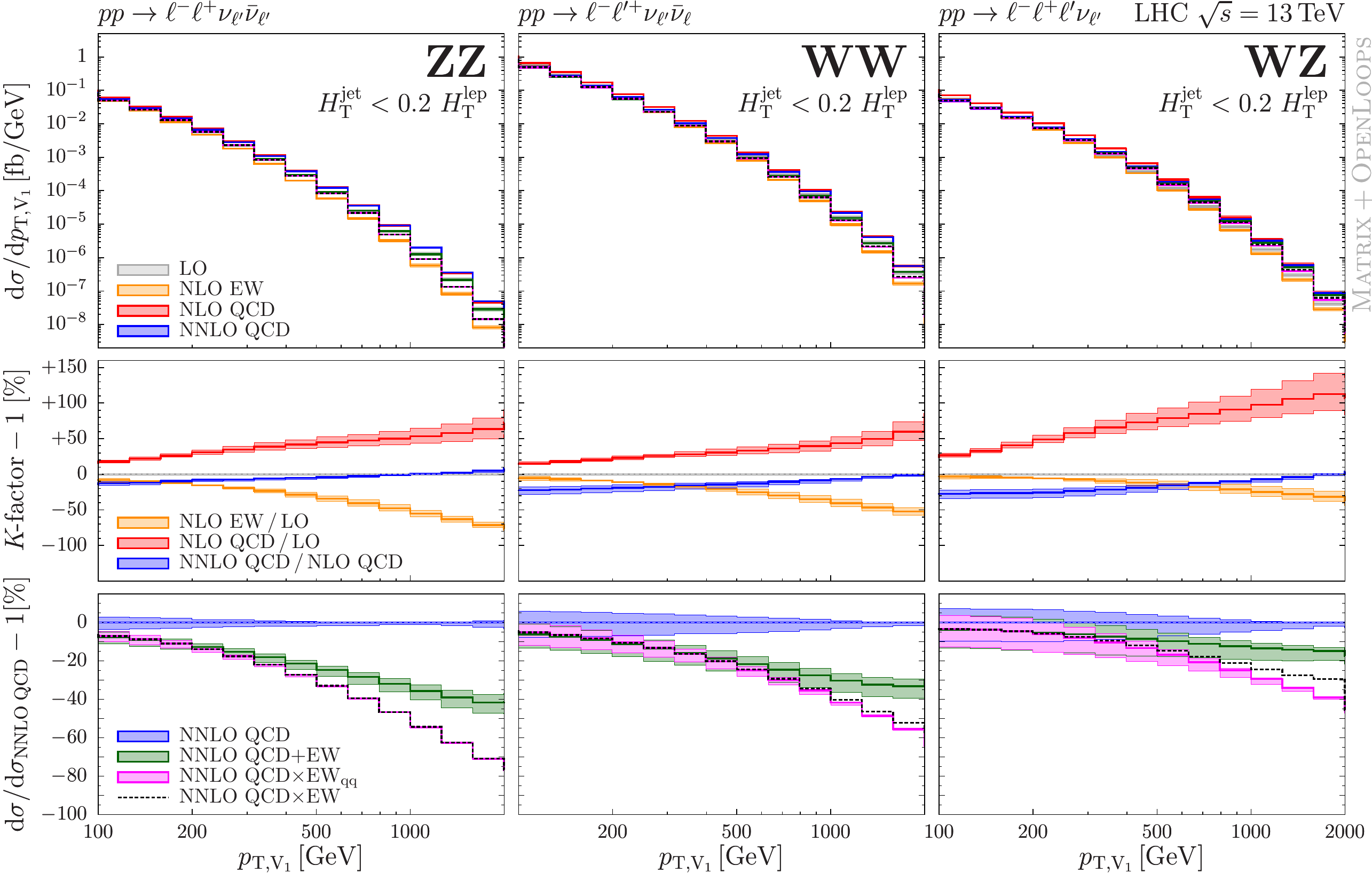}
\caption{
Distribution in the transverse momentum of the harder reconstructed vector
boson for the processes \refwoeq{proc:zz}--\refwoeq{proc:wz} at 13\,TeV. 
Same plots as in \fig{fig:inc_pTV1}, 
but in addition to the baseline cuts also the jet veto~\refwoeq{cuts:jetveto}
is applied. Note that here and in the following figures 
the corrections are given in percent.
\label{fig:veto_pTV1}
}
\end{figure*}

For the distribution in $p_{\rT,V_2}$ (not shown) the impact of the jet veto
turns out to be even milder than in the case of $m_{VV}$.  In contrast, in
the $p_{\rT,V_1}$ distribution,\footnote{The quantitative effects observed in the tail of
this distribution are documented in more detail in appendix~\ref{app:xsections}.}
shown in \reffi{fig:veto_pTV1}, the jet veto leads to a drastic suppression
of the giant $K$-factor effects observed in \reffi{fig:inc_pTV1}.
At 100\,GeV the NLO QCD corrections are reduced to \percentrange{20}{30}, and 
at large $\pT$ they grow much slower than in the 
inclusive case, reaching only \percent{50} in \zz and \ww production 
and \percent{100} in \wz production. This corresponds to a reduction 
by a factor of ten to twenty as compared to the inclusive case.
The moderate size of NNLO QCD corrections
(\percentrange{10}{20}) and the smallness of NNLO scale uncertainties
indicate that, in spite of the huge suppression of the \hardVj 
contributions induced by the jet veto,
the perturbative expansion of the \hardVV process remains 
free from large logarithmic QCD contributions beyond NLO.
In the presence of the jet veto also the huge positive EW corrections 
(see \reffi{fig:inc_pTV1})
are strongly suppressed. Hence, all three diboson processes feature large negative
corrections of Sudakov type that are 
qualitatively and quantitatively similar to
those observed in the tail of the $p_{\rT,V_2}$ distribution  without jet
veto. This is a further demonstration that, in the presence of the jet veto,
the distribution at hand is dominated by \hardVV production with 
\mbox{$p_{\rT,V_2}\sim p_{\rT,V_1}$}.
As a result of the jet veto, all combinations of QCD and EW
corrections yield large negative EW corrections independently of the
employed prescription. 
However, the sizeable NLO QCD effects in the tails
lead to non-negligible differences
between the additive and multiplicative combinations
(from a few percent at 500\,GeV to \percentrange{15}{30} at 2\,TeV). 
Nevertheless, given that the full NNLO QCD prediction is dominated by
\hardVV contributions, the multiplicative combinations can be regarded 
as realistic approximations of mixed QCD--EW corrections, and their differences with respect to
the additive combination is expected to largely overestimate $\ord(\alphaS\alpha)$ uncertainties.

\afterpage{
\begin{figure*}[t]
\centering
\includegraphics[width=\textwidth]{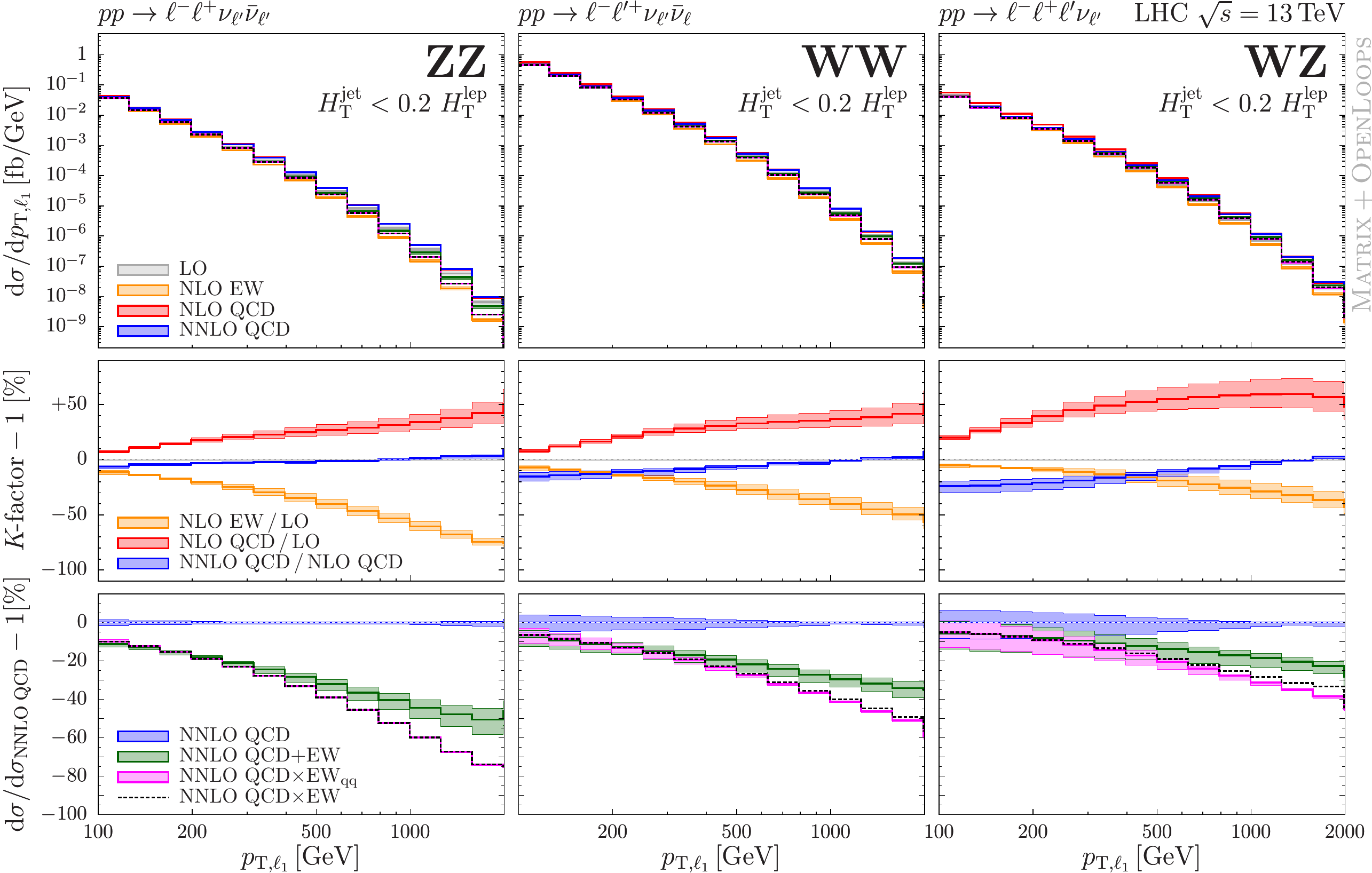}
\caption{
Distribution in the transverse momentum of the hardest charged lepton 
for the processes \refwoeq{proc:zz}--\refwoeq{proc:wz} at 13\,TeV. 
Same plots as in \fig{fig:inc_pTl1}, 
but in addition to the baseline cuts also the jet veto~\refwoeq{cuts:jetveto}
is applied.
\label{fig:veto_pTl1}
}
\vspace{1cm}
\centering
\includegraphics[width=\textwidth]{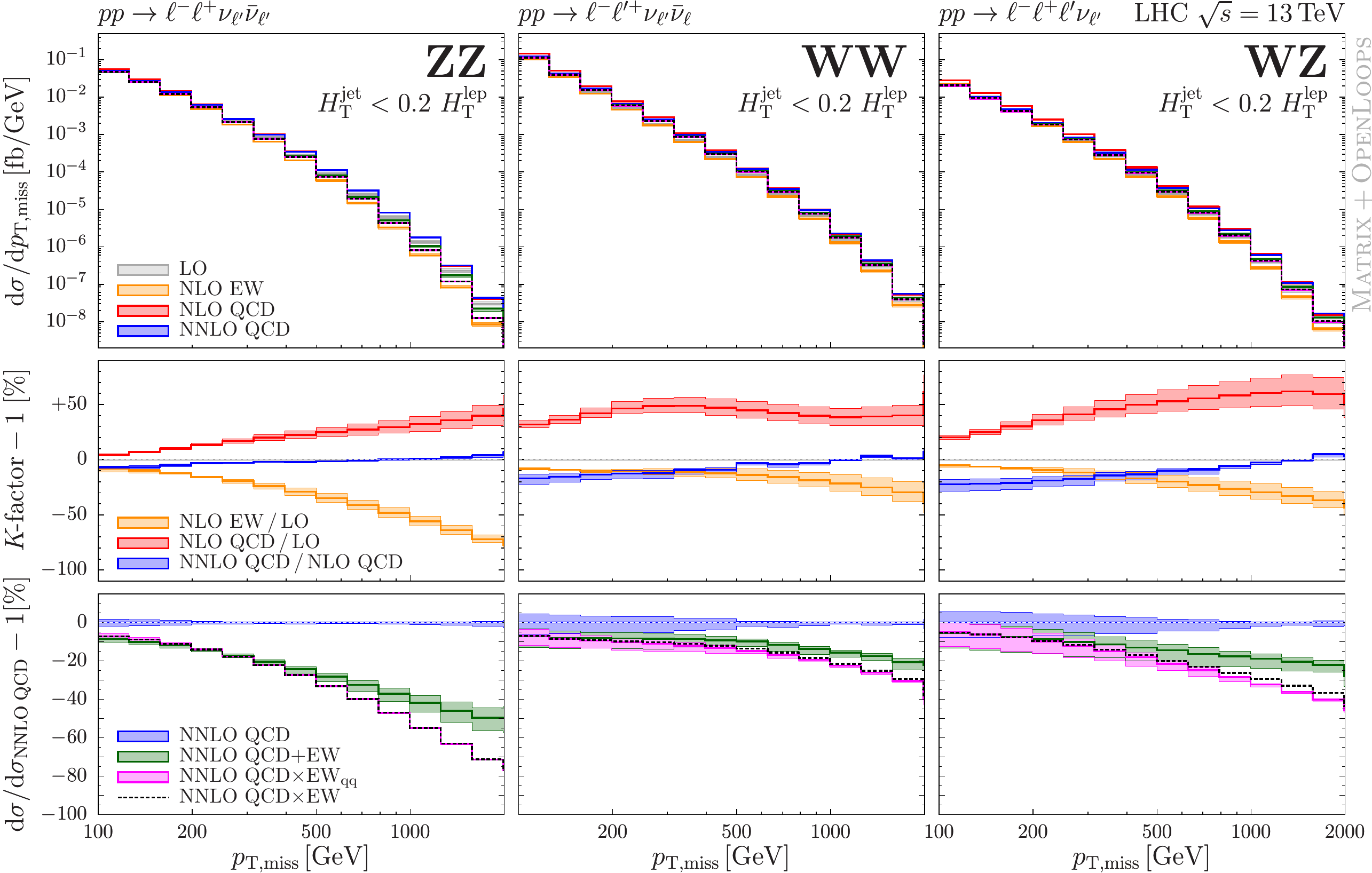}
\caption{
Distribution in the missing transverse momentum
for the processes \refwoeq{proc:zz}--\refwoeq{proc:wz} at 13\,TeV. 
Same plots as in \fig{fig:inc_MET}, 
but in addition to the baseline cuts also the jet veto~\refwoeq{cuts:jetveto}
is applied.
\label{fig:veto_MET}
}
\end{figure*}
\clearpage}

The jet veto affects the QCD and EW corrections to the
distribution in the $\pT$ of the hardest charged lepton, shown  
in \reffi{fig:veto_pTl1}, in a qualitatively very similar way as the 
distribution in $p_{\rT,V_1}$. 
This holds also for the distribution in the missing transverse momentum,
shown in  \reffi{fig:veto_MET}, where the 
giant QCD
$K$-factor due to the off-shell suppression of the LO \ww cross section 
(see \reffi{fig:inc_MET}) is reduced by up to a factor of twenty. Moreover, the
positive EW contributions from the \mbox{$\gamma q\to WV q$} channels 
disappear, and 
the EW corrections to all three diboson processes 
feature the usual Sudakov behaviour as expected for \hardVV production.

In general, the above observations support the application of selection cuts
that favour \hardVV topologies while suppressing 
\hardVj topologies and related giant $K$-factors
at high $\pT$.
This can be achieved either via an
appropriate jet veto, as illustrated here, or via a hardness requirement on
the subleading vector boson.
Besides reducing theoretical uncertainties from 
mixed QCD--EW higher-order effects, 
the suppression of \hardVj contributions
has the advantage of increasing the 
sensitivity of experimental measurements to 
genuine \hardVV topologies 
at high $\pT$, and thus to  
BSM effects related to anomalous gauge couplings.

\section{Summary and conclusions}
\label{sec:summa}

In this paper we have presented the first combination of NNLO QCD and NLO EW corrections for \wpwm, \wpmz and \zz 
production processes in their leptonic decay modes at the LHC.
Our calculation describes all the final states 
with two, three and four charged leptons. Spin correlations, interferences and off-shell effects are included throughout.

We have presented a detailed study of higher-order QCD and EW effects for three representative
leptonic channels, one for each of the \wpwm, \wpmz and \zz production modes.
The inclusion of QCD and EW corrections is 
crucial in order to reach percent-level precision for important kinematical distributions, 
as demanded by the experimental analyses.
The combination of QCD and EW corrections can be achieved in different ways.
We have compared 
an additive prescription (\NNLO \QCDpEW) and 
two multiplicative prescriptions where the factorisation of the EW corrections is complete (\NNLO \QCDtEW) or 
restricted to the $q\bar q$ channel (\NNLO \QCDtEWqq).
The differences between those prescriptions can be exploited to 
gain insights into higher-order QCD--EW mixed effects 
of $\ord(\alphaS\alpha)$ and related theoretical uncertainties.

The high-energy tails of various distributions turn out to be dominated by
the production of a hard vector boson in association with a hard jet and a
soft vector boson.  In these so-called \hardVj regions, QCD and EW
corrections behave in a completely different way as compared to regions
dominated by hard diboson production (\hardVV regions).
Thus theoretical uncertainties 
need to be assessed with different prescriptions for these
two kinds of regions.

The regions dominated by \hardVj production are characterised by 
the presence of giant QCD $K$-factors together with large negative EW Sudakov logarithms and possibly positive $\gamma$-induced corrections.  The interplay of these large $\ord(\alphaS)$ and
$\ord(\alpha)$ effects results in sizeable QCD--EW uncertainties
of $\ord(\alphaS\alpha)$ that cannot be controlled 
through a naive factorisation of the QCD and EW corrections.
In particular, the EW Sudakov corrections
to \mbox{$q\bar q\to VV$} production cannot be factorised since they are
not applicable to the 
\hardVj topologies that dominate the (N)NLO QCD cross section.
For these reasons, 
for observables dominated by \hardVj production, 
the band between the
$\NNLO\,\QCDpEW$ and $\NNLO\,\QCDtEWqq$ combinations
can be regarded as $\ord(\alphaS\alpha)$ uncertainty, and 
we suggest to use their average as nominal prediction.
Further reduction of these uncertainties requires the inclusion of EW
corrections to \mbox{$pp\to VVj$} production. The merging with
\mbox{$pp\to VV$} retaining \NNLO\,\QCDpEW accuracy is left to future work.

Giant $K$-factors and related $\ord(\alphaS\alpha)$ uncertainties 
are encountered only in rather inclusive observables at high $p_\rT$. Moreover, their 
appearance can be avoided by means of acceptance cuts that select 
\hardVV regions while suppressing \hardVj production.  This can be achieved through direct requirements on the
hardness of the two vector bosons or, as studied in this paper, through a
relatively mild jet veto.
Such phase space restrictions can also enhance the sensitivity of searches for anomalous gauge couplings at large energies.

In general, irrespectively of the presence of the jet veto,
in regions dominated by \hardVV production the perturbative QCD expansion is well behaved, 
and the large EW Sudakov logarithms in the tails of distributions 
are expected to factorise.
In those cases, the factorised prescriptions can be regarded 
as superior with respect to the additive one, and
the \NNLO \QCDtEWqq combination can be used as nominal prediction.
The difference between the two factorised prescriptions
can be regarded as uncertainty and is typically rather small,
while the difference between additive and multiplicative combinations
can largely overestimate $\ord(\alphaS\alpha)$ uncertainties.
A thorough estimate of $\ord(\alphaS\alpha)$
uncertainties for the \hardVV regions requires more detailed studies along the lines of
\citere{Lindert:2017olm} and is left to future work.

The NNLO QCD and NLO EW calculations presented in this paper have been carried out 
within the \MatrixOpenLoops framework, and their implementation paves the way to achieve 
the same level of accuracy for the hadronic production of an arbitrary colourless final state 
 for which the required two-loop QCD amplitudes are available. 
Theory predictions at NNLO QCD+NLO EW accuracy for the full set of massive diboson processes 
in the channels with two, three and four final-state charged leptons will be supported 
in the next release of \Matrix{}. Such predictions can be instrumental in boosting 
the sensitivity of a wide range of precision measurements and new-physics searches 
at the LHC and future $pp$ colliders.

\subsection*{Acknowledgements}
We would like to thank the {\sc NNPDF} collaboration, and in particular Juan Rojo and Maria Ubiali, for 
providing a suitable four-flavour NNLO PDF set including photon densities based on the LUXqed methodology. 
This work is supported in part
by the Swiss National Science Foundation (SNF) under contracts
BSCGI0-157722 and 200020-169041. The work of SK is supported by the ERC Starting Grant 714788 REINVENT. 
JL is supported by the STFC Rutherford Fellowship ST/S005048/1.
The work of MW was supported by the ERC Consolidator Grant 614577 HICCUP.

\appendix

\section{Cross sections in tails of $\pT$ distributions of the harder vector boson}
\label{app:xsections}

The behaviour of QCD and EW corrections observed in the tails of the 
$\pT$ distributions of the harder vector boson in figures~\ref
{fig:inc_pTV1} and \ref{fig:veto_pTV1}
is documented in more detail
in \refta{tab:cutHTjet_pTVleadM}, where
we present \zz, \ww and \wz cross sections 
with the same baseline cuts as in 
\refta{tab:baseline_inclusive} with an
additional cut \mbox{$p_{\mathrm{T,V_1}}>1\,\mathrm{TeV}$}.
Results are presented both for the case of inclusive QCD radiation and in the presence of 
the jet veto \refwoeq{cuts:jetveto}.

\renewcommand\arraystretch{1.4}
\begin{table}[t]
\begin{center}
\input{XStable_pTV1_M.tex}
\end{center}
\vspace*{-2ex}
\caption{Integrated cross sections for the processes \refwoeq{proc:zz}--\refwoeq{proc:wz}
at \mbox{$\sqrt{s}=13$\,TeV}. The baseline cuts~\refwoeq{cuts:pt}--\refwoeq{cuts:invm} are
supplemented by an additional cut on the transverse momentum of the leading
vector boson, \mbox{$p_{\mathrm{T,V_1}}>1\,\mathrm{TeV}$}.
Cross sections in the upper part of the table are inclusive with respect to QCD
radiation, while the lower part of the table reports corresponding results 
in the presence of the jet veto~\refwoeq{cuts:jetveto}.
The quoted uncertainties correspond to seven-point scale
variations. Absolute cross sections are reported in attobarn 
(\mbox{$1$\,ab $=10^{-3}$\,fb}).\\
}
\label{tab:cutHTjet_pTVleadM}
\end{table}

\clearpage

\bibliographystyle{JHEP}
\bibliography{4lqcdew}

\end{document}

%% file: defs.tex
\definecolor{darkgreen}{rgb}{0.0, 0.5, 0.0}

\newcommand{\GF}{{G_\mu}}

\newcommand{\LO}{\text{LO}\xspace}
\newcommand{\NLO}{\text{NLO}\xspace}
\newcommand{\NNLO}{\text{NNLO}\xspace}

\newcommand{\NNLL}{\text{NNLL}\xspace}

\newcommand{\QCD}{\text{QCD}\xspace}

\newcommand{\EW}{\text{EW}\xspace}

\newcommand{\EWqq}{\ensuremath{\EW_\text{qq}}\xspace}

\newcommand{\QCDpEW}{\text{\QCD{}+\EW}\xspace}

\newcommand{\QCDtEW}{\ensuremath{\text{\QCD{}\ensuremath{\times}\EW}}\xspace}
\newcommand{\QCDtEWqq}{\ensuremath{\text{\QCD{}\ensuremath{\times}\EWqq}}\xspace}

\newcommand{\Munich}{{\rmfamily \scshape Munich}\xspace}
\newcommand{\Matrix}{{\rmfamily \scshape Matrix}\xspace}
\newcommand{\MatrixOpenLoops}{{\rmfamily \scshape Matrix+OpenLoops}\xspace}
\newcommand{\Sherpa}{{\rmfamily\scshape Sherpa}\xspace}

\newcommand{\OpenLoops}{{\rmfamily\scshape OpenLoops}\xspace}

\newcommand{\RecolaTwo}{{\rmfamily\scshape Recola2}\xspace}

\newcommand{\Pa}{\ensuremath{\gamma}\xspace}
\newcommand{\Pt}{\ensuremath{t}\xspace}
\newcommand{\Pb}{\ensuremath{b}\xspace}

\newcommand{\PW}{\ensuremath{W}\xspace}

\newcommand{\PZ}{\ensuremath{Z}\xspace}
\newcommand{\PH}{\ensuremath{H}\xspace}

\newcommand{\wpwm}{\ensuremath{\PW^+\PW^-}\xspace}
\newcommand{\wpmz}{\ensuremath{\PW^\pm\PZ}\xspace}
\newcommand{\ww}{\ensuremath{\PW\PW}\xspace}
\newcommand{\zz}{\ensuremath{\PZ\PZ}\xspace}
\newcommand{\wz}{\ensuremath{\PW\PZ}\xspace}
\newcommand{\wpz}{\ensuremath{\PW^+\PZ}\xspace}
\newcommand{\wmz}{\ensuremath{\PW^-\PZ}\xspace}

\newcommand{\shortequal}{\ensuremath{\!\!\!=\!\!\!}}
\newcommand{\rsh}{\ensuremath{r_{21}}\xspace}

\newcommand{\hardVj}{hard-$Vj$\xspace}
\newcommand{\hardVV}{hard-$VV$\xspace}

\newcommand{\sig}[2]{\rd\sigma_{#1}^{#2}}
\newcommand{\delt}[2]{\delta_{#1}^{#2}}

\newcommand{\ri}{\mathrm{i}}
\newcommand{\rF}{\mathrm{F}}
\newcommand{\rR}{\mathrm{R}}

\newcommand{\rT}{\mathrm{T}}
\newcommand{\rd}{\mathrm{d}}

\newcommand{\rS}{\mathrm{S}}
\newcommand{\rw}{\mathrm{w}}
\newcommand{\MW}{M_\PW}
\newcommand{\MZ}{M_\PZ}
\newcommand{\MH}{M_\PH}
\newcommand{\Mt}{m_{\Pt}}
\newcommand{\Mb}{m_{\Pb}}

\newcommand{\GW}{\Gamma_{\PW}}
\newcommand{\GZ}{\Gamma_{\PZ}}
\newcommand{\GH}{\Gamma_{\PH}}
\newcommand{\Gb}{\Gamma_{\Pb}}
\newcommand{\Gt}{\Gamma_{\Pt}}

\newcommand{\mur}{\mu_{\rR}}
\newcommand{\muf}{\mu_{\rF}}
\newcommand{\murf}{\mu_{\rR,\rF}}

\def\xiveto{\xi_{\mathrm{veto}}}

\newcommand{\MeV}{\text{MeV}\xspace}
\newcommand{\GeV}{\text{GeV}\xspace}
\newcommand{\TeV}{\text{TeV}\xspace}

\newcommand{\alphaS}{\alpha_{\rS}}
\newcommand{\alphaW}{\alpha_{\rw}}

\newcommand{\ord}{\mathcal{O}}

\newcommand{\HTl}{H_{\mathrm{T}}^{\mathrm{lep}}}
\newcommand{\HTjet}{H_{\mathrm{T}}^{\mathrm{jet}}}

\newcommand{\pT}{\ensuremath{p_{\mathrm{T}}}\xspace}

\newcommand{\missingPT}{\ensuremath{p_{\mathrm{T,miss}}}\xspace}

\newcommand{\beqar}{\begin{eqnarray}}
\newcommand{\eeqar}{\end{eqnarray}}
\newcommand{\beq}{\begin{equation}}
\newcommand{\eeq}{\end{equation}}
\newcommand{\bsp}{\protect\begin{split}}
\newcommand{\esp}{\protect\end{split}}
\newcommand{\bit}{\begin{itemize}}
\newcommand{\eit}{\end{itemize}}

\newlength{\unitcharwidth}
\newcolumntype{C}{>{\centering\arraybackslash}p{0.165\textwidth}}

\newcommand{\refwoeq}[1]{\eqref{#1}}
\newcommand{\refeq}[1]{\mbox{eq.\ \eqref{#1}}}
\newcommand{\refeqs}[2]{\mbox{eqs.\ \eqref{#1}--\eqref{#2}}}
\newcommand{\reffi}[1]{\mbox{figure~\ref{#1}}}
\newcommand{\reffis}[2]{\mbox{figures~\ref{#1}--\ref{#2}}}
\newcommand{\refta}[1]{\mbox{table~\ref{#1}}}

\newcommand{\refse}[1]{\mbox{section~\ref{#1}}}
\newcommand{\refses}[2]{\mbox{sections~\ref{#1}--\ref{#2}}}

\newcommand{\citere}[1]{\mbox{ref.~\cite{#1}}}
\newcommand{\citeres}[1]{\mbox{refs.~\cite{#1}}}
\newcommand{\fig}{\reffi}
\newcommand{\sct}{\refse}
\newcommand{\eqn}{\refeq}

\newcommand{\ie}{i.e.\ }
\newcommand{\eg}{e.g.\ }

\newcommand{\as}{\alphaS}

\newcommand{\Rrec}{R_{\mathrm{rec}}}

\newcommand{\percentrange}[2]{\mbox{$#1$--$#2$\%}}
\newcommand{\percent}[1]{\mbox{$#1$\%}}

%% file: XStable_baseline.tex
\begin{tabular}{l|r|r|r}
  &	\multicolumn{1}{c|}{\zz}	&	\multicolumn{1}{c|}{\ww}	&	\multicolumn{1}{c}{\wz}	\\ \hline\hline 
		\multicolumn{1}{c|}{baseline cuts}	&	\multicolumn{1}{c|}{$pp\to\ell^-\ell^+\nu_{\ell'}\bar\nu_{\ell'}$}	&	\multicolumn{1}{c|}{$pp\to\ell^-\ell'^+\nu_{\ell'}\bar\nu_{\ell}$}	&	\multicolumn{1}{c}{$pp\to\ell^-\ell^+\ell'\nu_{\ell'}$}	\\ \hline\hline 
	                                         $\sigma_{\mathrm{LO}}$\,$[\mathrm{fb}]$	&	     $   18.9417(4)^{     +4.2 \%}_{     -5.2 \%}$	&	     $   240.221(5)^{     +5.1 \%}_{     -6.2 \%}$	&	     $   21.8960(3)^{     +4.2 \%}_{     -5.3 \%}$	\\
	                                    $\sigma_{\mathrm{NLO\,EW}}$\,$[\mathrm{fb}]$	&	     $   17.7713(4)^{     +4.3 \%}_{     -5.4 \%}$	&	     $   235.118(5)^{     +5.1 \%}_{     -6.1 \%}$	&	     $   21.1849(4)^{     +4.3 \%}_{     -5.4 \%}$	\\
	                                   $\sigma_{\mathrm{NLO\,QCD}}$\,$[\mathrm{fb}]$	&	     $    25.690(1)^{     +2.9 \%}_{     -2.4 \%}$	&	     $    370.35(1)^{     +4.2 \%}_{     -3.3 \%}$	&	     $    38.138(1)^{     +4.7 \%}_{     -3.8 \%}$	\\
	                                  $\sigma_{\mathrm{NNLO\,QCD}}$\,$[\mathrm{fb}]$	&	     $     29.63(2)^{     +3.0 \%}_{     -2.6 \%}$	&	     $     424.6(3)^{     +3.0 \%}_{     -2.7 \%}$	&	     $     42.28(2)^{     +2.3 \%}_{     -2.1 \%}$	\\
	                               $\sigma_{\mathrm{NNLO\,QCD+EW}}$\,$[\mathrm{fb}]$	&	     $     28.46(2)^{     +3.3 \%}_{     -2.7 \%}$	&	     $     419.5(3)^{     +3.0 \%}_{     -2.8 \%}$	&	     $     41.57(2)^{     +2.4 \%}_{     -2.2 \%}$	\\
	                         $\sigma_{\mathrm{NNLO\,QCD\times EW}}$\,$[\mathrm{fb}]$	&	     $     27.92(2)^{     +3.1 \%}_{     -2.6 \%}$	&	     $     416.0(3)^{     +3.0 \%}_{     -2.7 \%}$	&	     $     40.90(2)^{     +2.2 \%}_{     -2.1 \%}$	\\
	                    $\sigma_{\mathrm{NNLO\,QCD\times EW_{qq}}}$\,$[\mathrm{fb}]$	&	     $     27.92(2)^{     +3.1 \%}_{     -2.6 \%}$	&	     $     413.5(3)^{     +3.0 \%}_{     -2.7 \%}$	&	     $     40.53(2)^{     +2.1 \%}_{     -2.1 \%}$	\\[.5ex] \hline &&&\\[-4.5ex]
	                       $\frac{\sigma_{\mathrm{NLO\,EW}}}{\sigma_{\mathrm{LO}}}-1$	&	                                              $     -6.2 \%$	&	                                              $     -2.1 \%$	&	                                              $     -3.2 \%$	\\
	                      $\frac{\sigma_{\mathrm{NLO\,QCD}}}{\sigma_{\mathrm{LO}}}-1$	&	                                              $    +35.6 \%$	&	                                              $    +54.2 \%$	&	                                              $    +74.2 \%$	\\
	               $\frac{\sigma_{\mathrm{NNLO\,QCD}}}{\sigma_{\mathrm{NLO\,QCD}}}-1$	&	                                              $    +15.3 \%$	&	                                              $    +14.6 \%$	&	                                              $    +10.9 \%$	\\
	           $\frac{\sigma_{\mathrm{NNLO\,QCD+EW}}}{\sigma_{\mathrm{NNLO\,QCD}}}-1$	&	                                              $     -4.0 \%$	&	                                              $     -1.2 \%$	&	                                              $     -1.7 \%$	\\
	     $\frac{\sigma_{\mathrm{NNLO\,QCD\times EW}}}{\sigma_{\mathrm{NNLO\,QCD}}}-1$	&	                                              $     -5.7 \%$	&	                                              $     -2.0 \%$	&	                                              $     -3.2 \%$	\\
	$\frac{\sigma_{\mathrm{NNLO\,QCD\times EW_{qq}}}}{\sigma_{\mathrm{NNLO\,QCD}}}-1$	&	                                              $     -5.7 \%$	&	                                              $     -2.6 \%$	&	                                              $     -4.1 \%$	\\
\end{tabular}

%% file: XStable_pTV1_M.tex
\begin{tabular}{l|r|r|r}
	\multicolumn{1}{c|}{$p_{\mathrm{T,V_1}}>1\,\mathrm{TeV}$}	&	\multicolumn{1}{c|}{\zz}	&	\multicolumn{1}{c|}{\ww}	&	\multicolumn{1}{c}{\wz}	\\ \hline\hline 
		\multicolumn{1}{c|}{baseline cuts}	&	\multicolumn{1}{c|}{$pp\to\ell^-\ell^+\nu_{\ell'}\bar\nu_{\ell'}$}	&	\multicolumn{1}{c|}{$pp\to\ell^-\ell'^+\nu_{\ell'}\bar\nu_{\ell}$}	&	\multicolumn{1}{c}{$pp\to\ell^-\ell^+\ell'\nu_{\ell'}$}	\\ \hline\hline 
	                                         $\sigma_{\mathrm{LO}}$\,$[\mathrm{ab}]$	&	     $   0.41932(5)^{    +11.5 \%}_{     -9.7 \%}$	&	     $    5.2622(5)^{     +10. \%}_{     -8.4 \%}$	&	     $   0.56933(4)^{    +11.7 \%}_{     -9.8 \%}$	\\
	                                    $\sigma_{\mathrm{NLO\,EW}}$\,$[\mathrm{ab}]$	&	     $    0.1830(1)^{    +10.7 \%}_{     -9.1 \%}$	&	     $     5.429(3)^{     +6.4 \%}_{     -5.5 \%}$	&	     $    0.7875(2)^{     +6.9 \%}_{     -5.9 \%}$	\\
	                                   $\sigma_{\mathrm{NLO\,QCD}}$\,$[\mathrm{ab}]$	&	     $    1.5915(4)^{    +18.2 \%}_{    -14.0 \%}$	&	     $    35.950(6)^{    +21.1 \%}_{    -15.9 \%}$	&	     $    7.3451(8)^{    +23.8 \%}_{    -17.8 \%}$	\\
	                                  $\sigma_{\mathrm{NNLO\,QCD}}$\,$[\mathrm{ab}]$	&	     $     2.114(3)^{     +7.0 \%}_{     -7.7 \%}$	&	     $     46.39(4)^{     +6.1 \%}_{     -7.5 \%}$	&	     $     9.987(5)^{     +7.2 \%}_{     -8.6 \%}$	\\
	                               $\sigma_{\mathrm{NNLO\,QCD+EW}}$\,$[\mathrm{ab}]$	&	     $     1.877(3)^{     +6.4 \%}_{     -7.4 \%}$	&	     $     46.56(4)^{     +5.7 \%}_{     -7.2 \%}$	&	     $    10.205(5)^{     +6.9 \%}_{     -8.3 \%}$	\\
	                         $\sigma_{\mathrm{NNLO\,QCD\times EW}}$\,$[\mathrm{ab}]$	&	     $     0.923(3)^{     +6.5 \%}_{     -7.2 \%}$	&	     $     47.94(4)^{     +4.0 \%}_{     -4.5 \%}$	&	     $    14.422(5)^{     +4.6 \%}_{     -4.5 \%}$	\\
	                    $\sigma_{\mathrm{NNLO\,QCD\times EW_{qq}}}$\,$[\mathrm{ab}]$	&	     $     0.908(3)^{     +6.7 \%}_{     -7.3 \%}$	&	     $     26.66(4)^{     +5.5 \%}_{     -6.7 \%}$	&	     $     6.766(5)^{     +6.6 \%}_{     -7.9 \%}$	\\[.5ex] \hline &&&\\[-4.5ex]
	                       $\frac{\sigma_{\mathrm{NLO\,EW}}}{\sigma_{\mathrm{LO}}}-1$	&	                                              $    -56.4 \%$	&	                                              $     +3.2 \%$	&	                                              $    +38.3 \%$	\\
	                      $\frac{\sigma_{\mathrm{NLO\,QCD}}}{\sigma_{\mathrm{LO}}}-1$	&	                                              $   +279.5 \%$	&	                                              $   +583.2 \%$	&	                                              $  +1190.1 \%$	\\
	               $\frac{\sigma_{\mathrm{NNLO\,QCD}}}{\sigma_{\mathrm{NLO\,QCD}}}-1$	&	                                              $    +32.8 \%$	&	                                              $    +29.1 \%$	&	                                              $    +36.0 \%$	\\
	           $\frac{\sigma_{\mathrm{NNLO\,QCD+EW}}}{\sigma_{\mathrm{NNLO\,QCD}}}-1$	&	                                              $    -11.2 \%$	&	                                              $     +0.4 \%$	&	                                              $     +2.2 \%$	\\
	     $\frac{\sigma_{\mathrm{NNLO\,QCD\times EW}}}{\sigma_{\mathrm{NNLO\,QCD}}}-1$	&	                                              $    -56.3 \%$	&	                                              $     +3.3 \%$	&	                                              $    +44.4 \%$	\\
	$\frac{\sigma_{\mathrm{NNLO\,QCD\times EW_{qq}}}}{\sigma_{\mathrm{NNLO\,QCD}}}-1$	&	                                              $    -57.0 \%$	&	                                              $    -42.5 \%$	&	                                              $    -32.2 \%$	\\[1ex] \hline\hline
		\multicolumn{1}{c|}{$\HTjet< 0.2\HTl $}     &	\multicolumn{1}{c|}{$pp\to\ell^-\ell^+\nu_{\ell'}\bar\nu_{\ell'}$}	&	\multicolumn{1}{c|}{$pp\to\ell^-\ell'^+\nu_{\ell'}\bar\nu_{\ell}$}	&	\multicolumn{1}{c}{$pp\to\ell^-\ell^+\ell'\nu_{\ell'}$}	\\ \hline\hline 
	                                         $\sigma_{\mathrm{LO}}$\,$[\mathrm{ab}]$	&	     $   0.41932(5)^{    +11.5 \%}_{     -9.7 \%}$	&	     $    5.2622(5)^{     +10. \%}_{     -8.4 \%}$	&	     $   0.56933(4)^{    +11.7 \%}_{     -9.8 \%}$	\\
	                                    $\sigma_{\mathrm{NLO\,EW}}$\,$[\mathrm{ab}]$	&	     $    0.1798(1)^{    +10.9 \%}_{     -9.2 \%}$	&	     $     3.050(2)^{     +9.4 \%}_{     -8.0 \%}$	&	     $    0.4239(1)^{    +10.1 \%}_{     -8.5 \%}$	\\
	                                   $\sigma_{\mathrm{NLO\,QCD}}$\,$[\mathrm{ab}]$	&	     $    0.6482(2)^{     +7.5 \%}_{     -6.8 \%}$	&	     $     7.650(2)^{     +6.4 \%}_{     -5.8 \%}$	&	     $    1.1354(2)^{    +11.4 \%}_{     -9.5 \%}$	\\
	                                  $\sigma_{\mathrm{NNLO\,QCD}}$\,$[\mathrm{ab}]$	&	     $     0.656(2)^{     +0.2 \%}_{     -1.3 \%}$	&	     $      7.21(2)^{     +0.8 \%}_{     -2.9 \%}$	&	     $     1.067(2)^{     +0.6 \%}_{     -3.9 \%}$	\\
	                               $\sigma_{\mathrm{NNLO\,QCD+EW}}$\,$[\mathrm{ab}]$	&	     $     0.416(2)^{     +5.3 \%}_{     -7.5 \%}$	&	     $      5.00(2)^{     +5.1 \%}_{     -9.0 \%}$	&	     $     0.922(2)^{     +2.8 \%}_{     -7.1 \%}$	\\
	                         $\sigma_{\mathrm{NNLO\,QCD\times EW}}$\,$[\mathrm{ab}]$	&	     $     0.286(2)^{     +0.4 \%}_{     -1.3 \%}$	&	     $      4.20(2)^{     +1.1 \%}_{     -3.2 \%}$	&	     $     0.797(2)^{     +2.1 \%}_{     -5.3 \%}$	\\
	                    $\sigma_{\mathrm{NNLO\,QCD\times EW_{qq}}}$\,$[\mathrm{ab}]$	&	     $     0.286(2)^{     +0.4 \%}_{     -1.3 \%}$	&	     $      4.08(2)^{     +0.5 \%}_{     -2.1 \%}$	&	     $     0.740(2)^{     +0.7 \%}_{     -3.7 \%}$	\\[.5ex] \hline &&&\\[-4.5ex]
	                       $\frac{\sigma_{\mathrm{NLO\,EW}}}{\sigma_{\mathrm{LO}}}-1$	&	                                              $    -57.1 \%$	&	                                              $    -42.0 \%$	&	                                              $    -25.5 \%$	\\
	                      $\frac{\sigma_{\mathrm{NLO\,QCD}}}{\sigma_{\mathrm{LO}}}-1$	&	                                              $    +54.6 \%$	&	                                              $    +45.4 \%$	&	                                              $    +99.4 \%$	\\
	               $\frac{\sigma_{\mathrm{NNLO\,QCD}}}{\sigma_{\mathrm{NLO\,QCD}}}-1$	&	                                              $     +1.1 \%$	&	                                              $     -5.7 \%$	&	                                              $     -6.0 \%$	\\
	           $\frac{\sigma_{\mathrm{NNLO\,QCD+EW}}}{\sigma_{\mathrm{NNLO\,QCD}}}-1$	&	                                              $    -36.5 \%$	&	                                              $    -30.7 \%$	&	                                              $    -13.6 \%$	\\
	     $\frac{\sigma_{\mathrm{NNLO\,QCD\times EW}}}{\sigma_{\mathrm{NNLO\,QCD}}}-1$	&	                                              $    -56.4 \%$	&	                                              $    -41.8 \%$	&	                                              $    -25.3 \%$	\\
	$\frac{\sigma_{\mathrm{NNLO\,QCD\times EW_{qq}}}}{\sigma_{\mathrm{NNLO\,QCD}}}-1$	&	                                              $    -56.4 \%$	&	                                              $    -43.4 \%$	&	                                              $    -30.7 \%$	\\
\end{tabular}